\documentclass[usenatbib]{mnras}
\usepackage{multirow}
\usepackage{rotating}
\usepackage{colortbl}
\usepackage{color}
\usepackage[fleqn]{amsmath}
\usepackage{amssymb}
\usepackage{amsfonts}
\usepackage{verbatim}
\usepackage{scalefnt}
\usepackage[percent]{overpic}
\usepackage[T1]{fontenc} 
\usepackage{aecompl} 
\usepackage{ulem}

\def\msun{~{\rm M}_\odot}
\def\lsim{\mathrel{\rlap{\lower 3pt \hbox{$\sim$}} \raise 2.0pt \hbox{$<$}}}
\def\gsim{\mathrel{\rlap{\lower 3pt \hbox{$\sim$}} \raise 2.0pt \hbox{$>$}}}

\newcommand{\comments}[1]{} 

\title[Barred galaxies and feedback]{Barred galaxies in cosmological zoom-in simulations: \\the importance of feedback}

\author[T. Zana et al.]{Tommaso Zana,$^{1,2}$\thanks{E-mail: tzana@studenti.uninsubria.it} Pedro~R. Capelo,$^{3}$ Massimo Dotti,$^{4,5}$ Lucio Mayer,$^{3}$ \newauthor Alessandro Lupi,$^{6}$ Francesco Haardt,$^{2,5}$ Silvia Bonoli$^{7,8,9}$ and Sijing Shen$^{10}$\\
$^{1}$Scuola Normale Superiore, Piazza dei Cavalieri 7, IT-56126 Pisa, Italy\\
$^{2}$DiSAT, Universit\`a degli Studi dell'Insubria, Via Valleggio 11, IT-22100 Como, Italy\\
$^{3}$Center for Theoretical Astrophysics and Cosmology, Institute for Computational Science, University of Zurich,\\
Winterthurerstrasse 190, CH-8057 Z$\ddot{u}$rich, Switzerland\\
$^{4}$Dipartimento di Fisica G. Occhialini, Universit\`a di Milano-Bicocca, Piazza della Scienza 3, IT-20126 Milano, Italy\\
$^{5}$INFN, Sezione di Milano-Bicocca, Piazza della Scienza 3, IT-20126 Milano, Italy\\
$^{6}$Scuola Normale Superiore, Piazza dei Cavalieri 7, IT-56126 Pisa, Italy\\
$^{7}$Centro de Estudios de F{\'i}sica del Cosmos de Arag{\'o}n, plaza San Juan, 1 planta-2 ES-44001 Teruel, Spain\\
$^{8}$Donostia International Physics Center, Paseo Manuel de Lardizabal 4, ES-20018 Donostia-San Sebasti{\'a}n, Spain\\
$^{9}$IKERBASQUE, Basque Foundation for Science, ES-48013 Bilbao, Spain\\
$^{10}$Institute of Theoretical Astrophysics, University of Oslo, Postboks 1029, 0315 Oslo, Norway
}

\date{Accepted 2019 June 28. Received 2019 June 3; in original form 2018 October 16}

\pubyear{2019}

\begin{document}

\label{firstpage}

\pagerange{\pageref{firstpage}--\pageref{lastpage}}

\maketitle


\begin{abstract}
Bars are a key factor in the long-term evolution of spiral galaxies, in their unique role in redistributing angular momentum and transporting gas and stars on large scales. The Eris-suite simulations are cosmological zoom-in, $N$-body, smoothed-particle hydrodynamic simulations built to follow the formation and evolution of a Milky Way-sized galaxy across the build-up of the large scale structure. Here we analyse and describe the outcome of two particular simulations taken from the Eris suite -- ErisBH and Eris2k -- which mainly differ in the prescriptions employed for gas cooling, star formation, and feedback from supernovae and black holes. Our study shows that the enhanced effective feedback in Eris2k, due to the collective effect of the different micro-physics implementations, results in a galaxy which is less massive than its ErisBH counterpart till $z \sim 1$. However, when the stellar content is large enough so that global dynamical instabilities can be triggered, the galaxy in Eris2k develops a stronger and more extended bar with respect to ErisBH. We demonstrate that he structural properties and time evolution of the two bars are very different. Our results highlight the importance of accurate sub-grid prescriptions in cosmological zoom-in simulations of the process of galaxy formation and evolution, and the possible use of a statistical sample of barred galaxies to assess the strength of the stellar feedback.
\end{abstract}



\section{Introduction}\label{sec:introduction}

Since the seminal papers by \citet{Hubble_1936} and \citet{de_Vaucouleurs_1963}, bars have been considered major actors in the long-term evolution of spiral galaxies \citep{Kormendy_Kennicutt_2004, Kormendy_2013}.
Bars provide a deviation from an otherwise axisymmetric potential and, thus, allow both to redistribute angular momentum and to transport stellar and gaseous components on local and global scales \citep{Tremaine_Weinberg_1984, Lynden-Bell_Kalnajs_1972, Athanassoula_2003}.
Amongst the consequences driven by this migration process, the aging of the inner galactic environment \citep[e.g.][]{Cheung_et_al_2013, Gavazzi_et_al_2015, Consolandi_et_al_2016, Consolandi_et_al_2017, Khoperskov_etal_2018} and the formation of major sub-structures, such as the frequently observed boxy-peanut bulges \citep[e.g][]{Combes_et_al_1990, Kormendy_1993, Chung_Bureau_2004, Kormendy_1993, Athanassoula_2005, Athanassoula_2008}, or the characteristic star-forming rings \citep[e.g.][]{Buta_2004, Romero-Gomez_et_al_2007}, are intimately connected to the presence of a bar. 
For these reasons, the study of the properties of the bar and of its complex interplay with the host system has become a key factor in modern galaxy evolution theories.
When $N$-body techniques were first applied to stellar dynamical problems, many significant steps forward were taken on the way to investigating the non-axisymmetries formation and growth \citep{Miller_et_al_1970, Hohl_1971, Ostriker_Peebles_1973, Sellwood_Athanassoula_1986, Pfenniger_Friedli_1991} and their effects on the disc material \citep[e.g.][]{Sanders_Huntley_1976, Roberts_et_al_1979, Athanassoula_1992, Ho_et_al_1997, Martinet_Friedli_1997, Ho_et_al_1997, Lutticke_et_al_2000, Laurikainen_et_al_2004, Bureau_Athanassoula_2005, Jogee_et_al_2005, Kormendy_2013, Cheung_et_al_2013, Fanali_et_al_2015, Hakobyan_et_al_2016, Consolandi_et_al_2017}.
However, it is only with the recent improvements in spatial and mass resolution of numerical simulations and thanks to the development of state-of-the-art sub-grid recipes that the topic can be properly addressed in the cosmological context.

The theoretical study of barred galaxies in a fully cosmological framework has started only recently, following the first zoom-in simulations that produced realistic late-type galaxies.
Indeed, both the combination of high resolution and efficient stellar-feedback prescriptions are required in order to prevent the accumulation of central low angular momentum gas, allowing for the build-up of cold discs with flat rotation curves comparable with observations \citep{Navarro_Benz_1991}.

The majority of these recent works merely acknowledges the presence of a stellar bar within the galactic disc, since such structure does not represent the main subject of their analysis \citep[see, e.g.][]{Robertson_et_al_2004, Scannapieco_et_al_2009, Feldmann_et_al_2010, Brooks_et_al_2011, Bonoli_et_al_2016, Sokolowska_et_al_2017}.
However, there are also some studies specifically aimed to the analysis of the evolving non-axisymmetry \citep[e.g.][]{Romano-Diaz_et_al_2008, Kraljic_et_al_2012, Scannapieco_Athanassoula_2012, Goz_et_al_2015, Okamoto_et_al_2015, Spinoso_et_al_2017, Zana_et_al_2018a, Zana_et_al_2018b}.
Mass and spatial resolution significantly improved over the years and this allowed to follow the investigated bar properties in ever greater detail.
An alternative approach is offered by \citet{Algorry_et_al_2017} and \citet{Peschken_Lokas_2018}, who perform statistical studies taking advantage of the large-volume cosmological simulations EAGLE \citep{Crain_et_al_2015, Schaye_et_al_2015} and Illustris \citep{Vogelsberger_et_al_2014}, respectively. 
Nonetheless, the large sample of galaxies in these simulations has been achieved by implementing lower mass and time resolution, consequently hindering the study on sub-galactic scales.
A special mention has to be reserved for \citet{Governato_et_al_2007}, where the authors implement different sub-grid prescriptions to simulate three galaxy-sized haloes in hydro-cosmological runs with remarkable resolution -- a gravitational softening ranging from $0.3$ to $1$~kpc -- but still not optimal in order to properly characterise the evolving features of a sub-galactic structure. Even though the focus of their work was essentially the study of the formed galactic systems, the authors were the first to point out a direct effect of the supernova (SN) feedback over the development of a stellar bar, finding that the main effect of SN energy injection was to make the disc more stable against bar formation, by contributing to the formation of a lighter stellar disc which builds up more slowly over time.

The Eris-suite simulations (e.g. Eris, \citealt{Guedes_et_al_2011}; ErisLE, \citealt{Bird_et_al_2013}; ErisBH, \citealt{Bonoli_et_al_2016}; and Eris2k, \citealt{Sokolowska_et_al_2016,Sokolowska_et_al_2017}) succeeded in reproducing a set of realistic spiral galaxies in zoom-in cosmological volumes, thanks to a smart management of the feedback recipes, and provide a fruitful laboratory to explore numerous physical processes in a cosmological context. In particular, the runs ErisBH \citep{Bonoli_et_al_2016} and Eris2k \citep{Sokolowska_et_al_2016, Sokolowska_et_al_2017} -- described in Section~\ref{sec:numerical_setup} -- whilst sharing the same initial conditions, implement different unresolved stellar-physics prescriptions, leading to the formation of two realistic, although different, disc galaxies, both hosting kpc-scale stellar bars.

In this paper, we detail the differences between the main galaxies forming in the two above-mentioned cosmological runs: in the whole disc (Section~\ref{sec:global_scales}), in the formation and growth of their bars (Section~\ref{sec:local_scales}), and in the following deaths of these non-axisymmetric structures (Section~\ref{sec:local_scales2}). Section~\ref{sec:formation} presents a dynamical analysis that links the observed differences to the distinct stellar-physics sub-resolution prescriptions. Finally, we present our conclusions in Section~\ref{sec:conclusions}.


\section{Numerical setup}\label{sec:numerical_setup}

The two simulations analysed in this work -- Eris2k and ErisBH -- are part of the Eris suite, a family of cosmological zoom-in simulations built to follow the formation and evolution of a local Milky Way (MW)-sized galaxy and run with the $N$-body, smoothed-particle hydrodynamics code {\textsc{gasoline}} \citep[][]{Stadel_2001,Wadsley_et_al_2004}.

All simulations in the suite share the same cosmological parameters (from the Wilkinson Microwave Anisotropy Probe three-yr data: $\Omega_{\rm M} = 0.24$, $\Omega_{\Lambda} = 1 - \Omega_{\rm M}$, $\Omega_{\rm b} = 0.042$, $h = 0.73$, $n = 0.96$, and $\sigma_8 = 0.76$; \citealt{Spergel_et_al_2007}) and the same cosmological box of (90 comoving Mpc)$^3$, within which a low-resolution dark-matter (DM)-only simulation with $300^3$ DM particles is run from redshift $z = 90$ down to $z = 0$. After a target halo (with halo mass similar to that of the MW and a late quiet merging history, i.e. with no major mergers -- above a mass ratio of 0.1 -- after $z = 3$) is chosen, a zoom-in hydrodynamical simulation is performed with $1.3 \times 10^7$ DM particles and $1.3 \times 10^7$ gas particles within a Lagrangian sub-volume of (1 comoving Mpc)$^3$ around such a halo. The mass and spatial resolution in the high-resolution region are given by the mass of DM ($m_{\rm DM} = 9.8 \times 10^4 \msun$) and gas ($m_{\rm gas} = 2 \times 10^4$~M$_{\odot}$) particles, and by the gravitational softening of all particle species: $\epsilon = 1.2/(1+z)$~kpc for $90 \ge z > 9$ and 0.12~kpc for $z\le 9$.

The main differences within the suite lie in how (i) gas cooling, (ii) stellar models (including star formation -- SF -- and feedback from SNae), and (iii) black hole (BH) physics are implemented. In the following, we will focus on the distinction between ErisBH (run down to $z_{\rm end} = 0$) and Eris2k ($z_{\rm end} = 0.31$).\footnote{The Eris2k simulation had to be halted at $\sim$10~Gyr due to the exhaustion of computational resources.}

Both simulations include Compton cooling and primordial atomic non-equilibrium cooling in the presence of a redshift-dependent cosmic ionizing background (\citealt{Haardt_Madau_2012} in Eris2k and \citealt{Haardt_Madau_1996} in ErisBH). The modelling of metal cooling varies amongst simulations. In ErisBH, cold gas ($T_{\rm gas} < 10^4$~K) can cool from the de-excitation of fine structure and metastable lines (C, N, O, Fe, S, and Si), and is maintained in ionization equilibrium by a local cosmic ray flux \citep{Bromm_et_al_2001,Mashchenko_et_al_2008}. In Eris2k, a look-up table \citep[][]{Shen_et_al_2010,Shen_et_al_2013} is used, in which the cooling rates for the first 30 elements in the periodic table are pre-computed as a function of gas temperature ($10 \le T_{\rm gas} \le 10^9$~K), density ($10^{-9} \le n_{\rm H} \le 10^4$~cm$^{-3}$, where $n_{\rm H}$ is the hydrogen number density), and redshift ($0 \le z \le 15.1$), using {\textsc{cloudy}} \citep[][]{Ferland_et_al_1998} and assuming photo-ionization equilibrium (PIE) with the cosmic ionizing background. Cooling is stronger in Eris2k at all temperatures: for $T_{\rm gas} > 10^4$~K, the presence of metals can increase cooling rates even by orders of magnitude \citep[][]{Shen_et_al_2010}; for $T_{\rm gas} < 10^4$~K, it was shown that PIE cooling is enhanced with respect to  collisional ionization equilibrium cooling (what used in ErisBH, assuming cosmic rays are unimportant), mostly because PIE tables are computed under the assumption of an optically thin gas and therefore overestimate the true metal cooling rates of low-temperature gas \citep[][]{Bovino_et_al_2016, Capelo_et_al_2018}.

Gas particles denser than $\rho_{\rm SF}$, colder than $T_{\rm SF}$, and with a local gas overdensity $>$2.63 are allowed to form stars, i.e. stochastically converted into star particles so that ${\rm d}M_*/{\rm d}t = \epsilon_* M_{\rm gas}/t_{\rm dyn}$, where $M_{\rm gas}$ and $M_*$ are the mass of gas and stars involved in the SF event, respectively, $t_{\rm dyn} = (4\pi {\rm G} \rho_{\rm gas})^{-1/2}$ is the local dynamical time, $\rho_{\rm gas}$ is the gas density, $G$ is the gravitational constant, and $\epsilon_* = 0.1$ is the SF efficiency \citep[][]{Stinson_et_al_2006}. Stars form in a denser ($\rho_{\rm SF} = 10^2 \,m_{\rm H}$ versus $5 \,m_{\rm H}$~cm$^{-3}$, where $m_{\rm H}$ is the hydrogen mass) and colder ($T_{\rm SF} = 10^4$ versus $3 \times 10^4$~K) gas environment in Eris2k than in ErisBH.

At each SF event, the new stellar particles (each of mass $\sim$ $6 \times 10^3 \msun$) are a proxy for a stellar population with a given initial mass function (IMF). The different IMF implemented (\citealt{Kroupa_2001} in Eris2k and \citealt{Kroupa_et_al_1993} in ErisBH) translates into about three times more stars in the mass range 8--40~M$_{\odot}$ in Eris2k than in ErisBH, for a fixed stellar particle mass \citep[][]{Shen_et_al_2013,Sokolowska_et_al_2016}. Such difference is relevant because stars with mass within 
that range can explode as SNae, injecting mass, metals, and thermal energy into the surrounding gas, according to the ``blastwave model'' of \citet{Stinson_et_al_2006}. The energy from SNae ($E_{\rm SN} \equiv \epsilon_{\rm SN} \times 10^{51}$ erg per SN, with $\epsilon_{\rm SN} = 0.8$ and $\epsilon_{\rm SN} = 1$ for ErisBH and Eris2k, respectively) heats the surrounding gas particles, which are then allowed to cool radiatively, but only after a cooling shut-off time equal to the survival time of the hot low-density shell of the SN (\citealt{McKee_Ostriker_1977}; in ErisBH) or twice that (in Eris2k), in order to prevent the gas from quickly radiating away the SN energy because of the limited resolution.

In Eris2k, thermal energy and metals are turbulently diffused \citep[][]{Wadsley_et_al_2008, Shen_et_al_2010}, whereas in ErisBH this applies to the thermal energy only.
When such mechanism is implemented, the metallicity distribution in the interstellar medium, as a function of the density, becomes smoother: both the formation of low-density zones with high metallicity is prevented, and the total amount of metals inside the galactic halo rises.
The major consequence is that the cooling rate is further increased in the galaxy disc, and the SF process is then (slightly) favoured (see \citealt{Shen_et_al_2010} for further details).

With respect to Eris2k (and to all other simulations in the Eris suite), ErisBH includes additional prescriptions, in the form of seeding, accretion, feedback, and merging of BHs. BH seeds are inserted at the centre of a given halo when such a system is bound, is resolved by at least $10^5$ particles, has at least 10 gas particles with density $> 10^2 \,m_{\rm H}$~cm$^{-3}$, and does not already host a BH. The seeded BH has an initial mass proportional to the number of high-density gas particles. BHs are then allowed to accrete the surrounding gas according to the commonly used Bondi--Hoyle--Lyttleton formula \citep[][]{Hoyle_Lyttleton_1939,Bondi_Hoyle_1944,Bondi_1952}, with a maximum allowed accretion rate set by the \citet{Eddington_1916} limit. While accreting, BHs exert feedback by injecting onto the surrounding medium, in the form of thermal energy, 5 per cent of the radiated luminosity. Finally, two BHs can merge if they have short separations and low relative velocities \citep[][]{Bellovary_et_al_2010}.

We caution the reader that, when we talk about different feedback, we do not refer only to the SN efficiency parameter $\epsilon_{\rm SN}$, but rather to the ``effective feedback'' given by the cumulative effect of all sub-grid parameters.
On one hand, the combined unresolved-physics prescriptions and parameters in Eris2k with respect to ErisBH (enhanced cooling, larger SF density threshold, different IMF, increased SN energy, and longer cooling shut-off time) lead to a globally increased stellar effective feedback in the former simulation \citep[][]{Sokolowska_et_al_2016,Sokolowska_et_al_2017}. On the other hand, the implementation of BHs in ErisBH leads to a potentially boosted feedback effect in the central regions of the galaxy, although this strongly depends on the mass of the BH, which in ErisBH reaches $\sim$$2.6 \times 10^6 \msun$ at $z = 0$ \citep[][]{Bonoli_et_al_2016}.

Both the runs result in the formation of two barred MW-sized disc galaxies of stellar mass $\gsim 10^{10} \msun$, showing no ``classical'' bulge component. 
The total virial mass of ErisBH (at $z = 0$) is $8.2 \times 10^{11} \msun$ within a virial radius of $265$ physical\footnote{In the rest of this work, unless otherwise specified, we always report the physical lengths.} kpc, whereas the main galaxy in Eris2k (at $z = 0.31$) has a virial total mass and radius of $7.5 \times 10^{11} \msun$ and $211$~kpc, respectively.

The stellar surface density $\Sigma_*$  and its decomposition into sub-components are shown for the final snapshots of the two runs in Figure~\ref{fig:fit} (see Appendix~\ref{sec:fit} for the entire fitting procedure). 
The details of the global properties of the galaxies and of their sub-structures and their dependencies on the different baryonic-physics prescriptions will be the focus of the next two sections.

\begin{figure}
\vspace{-9pt}
\includegraphics[width=\columnwidth]{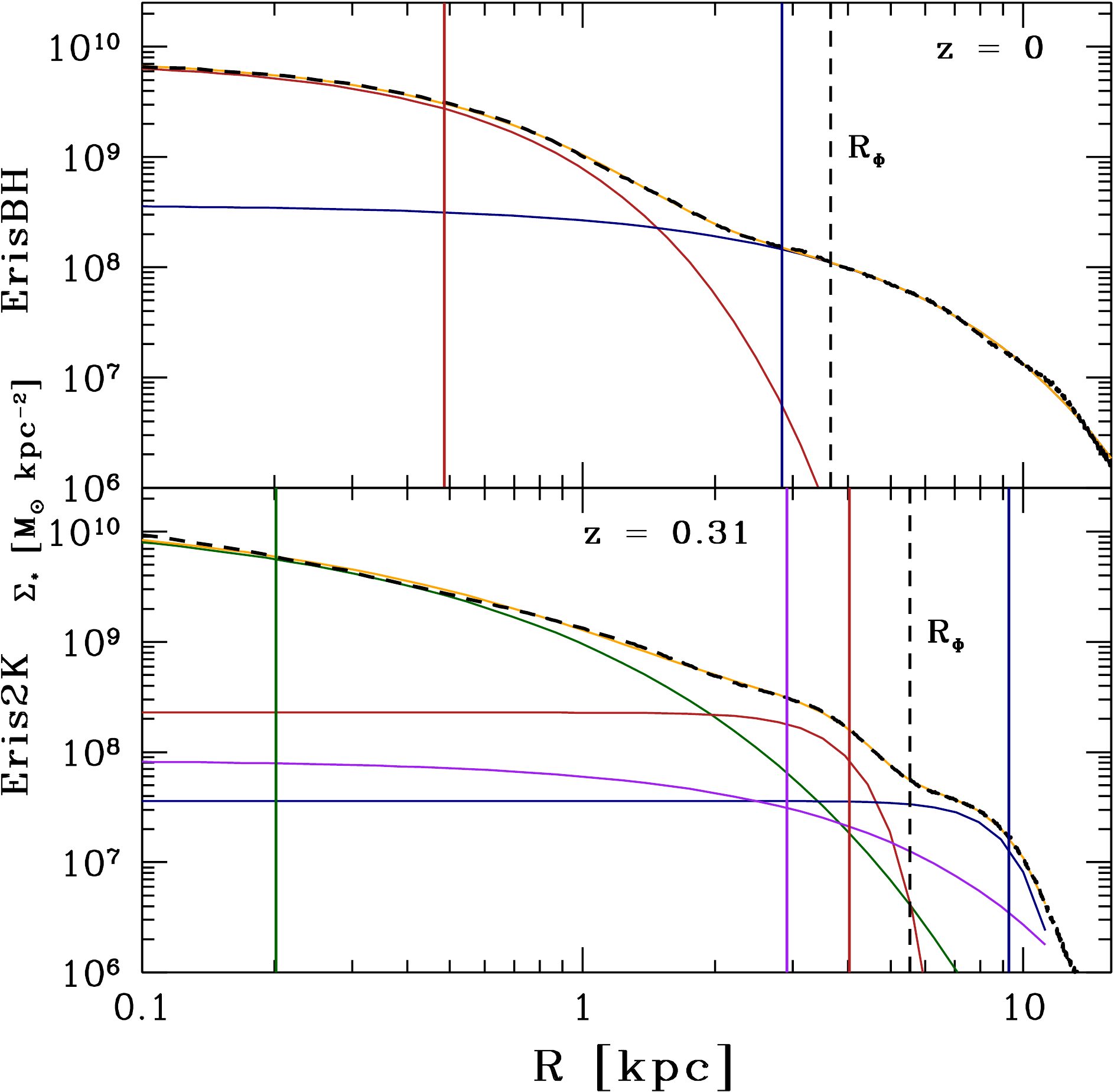}
\caption{Profile decomposition of the stellar surface density $\Sigma_*$ for the ErisBH (upper panel) and Eris2k (lower panel) runs at their last snapshot. The black lines mark the profiles of $\Sigma_*$ measured from the snapshots, whereas the yellow lines refer to the best fits obtained with two \citet{Sersic_1963,Sersic_1968} components for ErisBH and 3 S\'ersic components and an exponential profile for Eris2k. Both galaxies require a stellar disc component (blue curve) and a less extended component (red curve) associated with the stellar bar or, possibly, to its central part inflated into a boxy-peanut bulge in the case
of ErisBH (see Section~\ref{sec:local_scales2}). These two stellar systems are easily recognizable in the closeups of Figure~\ref{fig:proj}.
Eris2k shows a third, central, component associated to a recent burst of SF fuelled by bar-driven gas inflows (green curve; see Section~\ref{sec:local_scales2}) and a fourth component (magenta curve) to fit the background. The vertical solid lines mark the position of the scale radii of the corresponding fits with the same colour code, whereas the black dashed lines show the bar length ($R_{\Phi}$) defined in Section~\ref{sec:local_scales}.
See Appendix~\ref{sec:fit} for a description of the fitting method.}
\label{fig:fit}
\end{figure}


\section{Feedback effects on global scales: galaxy growth histories}\label{sec:global_scales}

\begin{figure}
\vspace{2pt}
\includegraphics[width=\columnwidth]{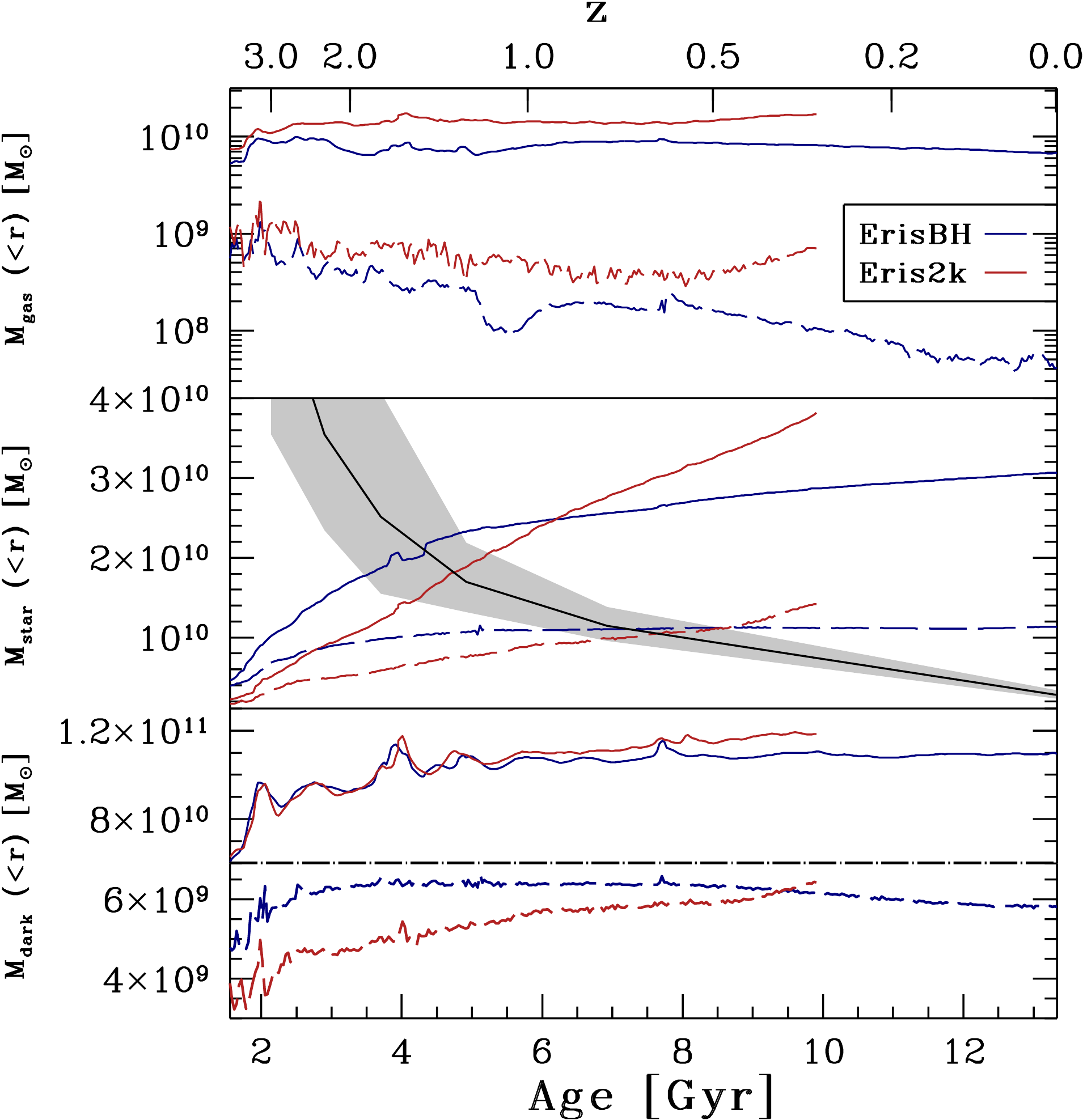}
\caption{From top to bottom: gas (first panel), stellar (second panel), and DM (third and fourth panels) masses enclosed within spheres of 20 (solid lines) and 2 (dashed lines) kpc radii. The blue lines refer to the ErisBH run,  whereas the red lines to Eris2k. The galaxy in the Eris2k simulation, from $z=3$ till the end of the run, approximately quadruples its stellar mass within both the analysed spheres, whereas the evolution is more modest in the ErisBH case. 
The peak in DM mass within 20 kpc at $z\sim1.5$ in both runs corresponds to the close interaction with a satellite.
The black solid line in the second panel refers to the knee mass in the relation between specific SF rate and stellar mass in \citeauthor{Gavazzi_et_al_2015} (\citeyear{Gavazzi_et_al_2015}; the errors are indicated by the shaded area), who proposed it as a threshold for the growth of stellar bars (see the discussion in Section~\ref{sec:conclusions}). 
}
\label{fig:mass}
\end{figure}

As discussed above, ErisBH and Eris2k share the same cosmological initial conditions and, as a consequence, are hosted in extremely similar large-scale DM haloes. Only the central regions of these DM haloes differ one from the other because of the unequal evolution of the baryons that dominate the central dynamics (see the two lowermost panels in Figure~\ref{fig:mass}). This is mostly due to the different implementations of the IMF in the two runs\footnote{The significant effect on SN feedback produced by different assumptions on the IMF has been recently addressed in cosmological zoom-in simulations \citep[see][]{Valentini_et_al_2019}.}, which result in a considerable disparity in the energy input into the interstellar medium via SN feedback. Indeed, the specific SN feedback energy input in new stars is $3.9 \times 10^{48}$ and $1.0 \times 10^{49} {\rm erg} \msun^{-1}$ in ErisBH and Eris2k, respectively\footnote{The quantities are calculated by integrating the high-mass tail of each IMF from 8 to 40~$\msun$ and considering the related efficiency $\epsilon_{\rm SN}$ \citep[see also][]{Shen_et_al_2013, Sokolowska_et_al_2016}.} (i.e. the SN energy per unit mass is more than 2.5 times higher in the Eris2k run).
Concurrently, the difference in the SF thresholds used in the two simulations produces only a second-order effect.
The minimum density $n_{\rm SF}$ has the greatest impact on the timing of SF, rather than on the resulting total mass of formed stars.
From a collapsing gas cloud of sufficient mass, a higher SF threshold would be still reached, though later in time. Indeed, some recent works showed that the total stellar mass is only minimally affected by the variation of $n_{\rm SF}$ \citep[see, e.g.][]{Lupi_et_al_2017}.
The enhanced effective feedback in Eris2k results in a larger fraction of the gas being preserved from forming stars and, thus, in a delayed stellar mass growth, as shown in the two uppermost panels of Figure~\ref{fig:mass}.
The steady increase of stellar mass produces, in turn, an increment of the DM component within the inner $20$~kpc (of about $10^{10} \msun$ with respect to ErisBH), caused by the consequent adiabatic contraction \citep[e.g.][]{Blumenthal_et_al_1986}.

Interestingly, the radial extent of the galaxies is less sensitive to the different amount of effective feedback. Due to the large (and varying) number of components needed to accurately fit $\Sigma_*$ in the two runs, and the uncertainty associated to the fitting of an inherently elongated structure (the bar) with an axisymmetric component (see Appendix~\ref{sec:fit}), we prefer not to estimate the disc extent directly from the scale length of the fitted disc component (see Figure~\ref{fig:fit} and Figure~\ref{fig:fit_2} for two examples of such decomposition). We decided instead to compute the \citet{Kron_1980} radius 
$R_{\rm K}$, 
i.e. a mass-averaged radius of the galaxy:\footnote{Note that the original formulation of the Kron radius weighs the radii using the stellar surface brightness, whereas here we are interested in the actual mass distribution. The two approaches are equivalent under the assumption of a $R$-independent mass-to-light ratio.}   

\begin{equation}\label{eq:kron}
R_{\rm K}(R)=\frac{\int_{0}^{R} \Sigma_*(x) x^2 dx}{\int_{0}^{R} \Sigma_*(x) x dx},
\end{equation}

\noindent evaluated at the radius $R$ where ${\rm d}R_{\rm K}/{\rm d}R<0.03$.\footnote{Such definition is needed to prevent the contamination by stars not associated to the main galaxy.}

\begin{figure*}
\vspace{-8pt}
\hspace{-4pt}
\includegraphics[width=0.94\textwidth]{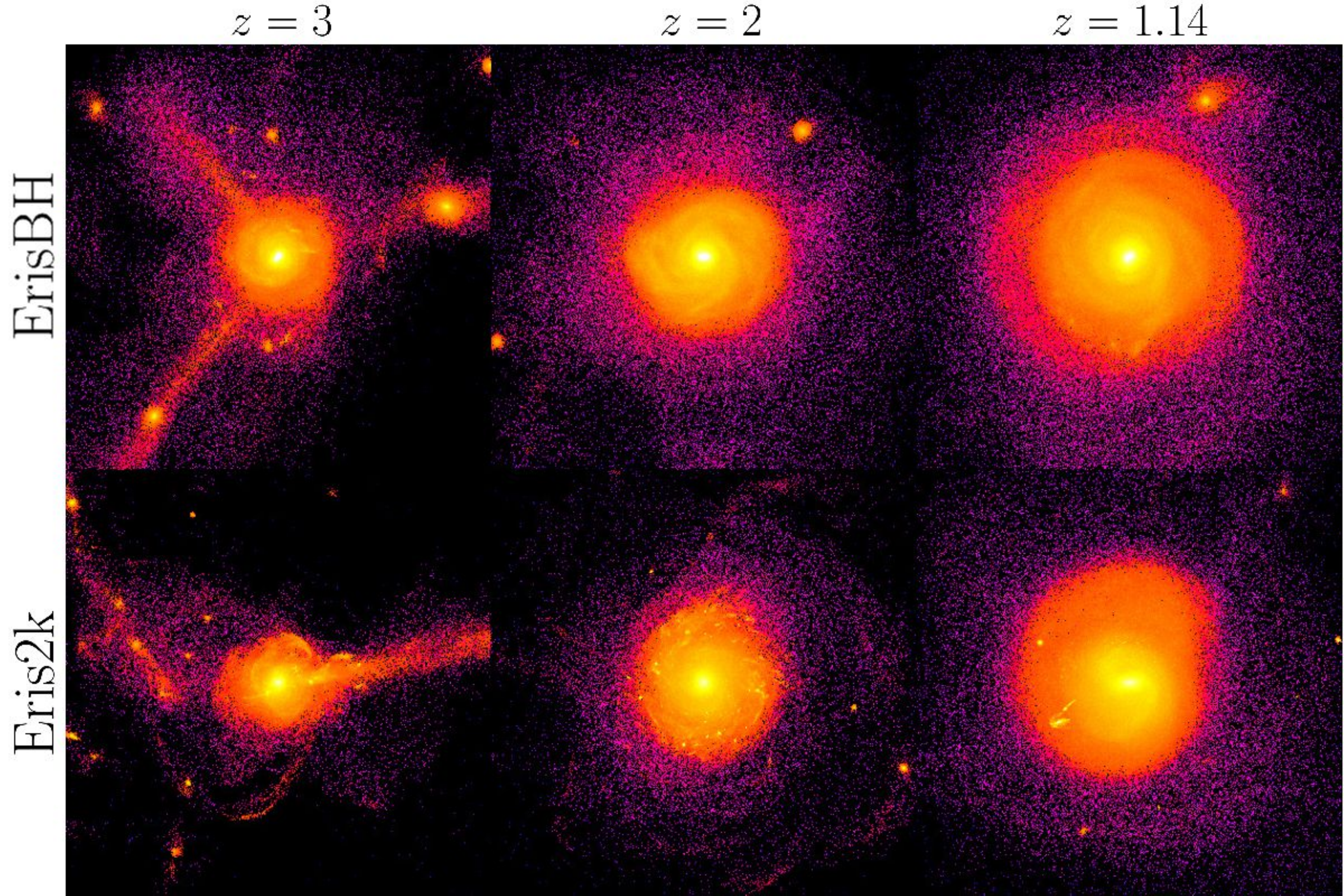}
\includegraphics[width=0.94\textwidth]{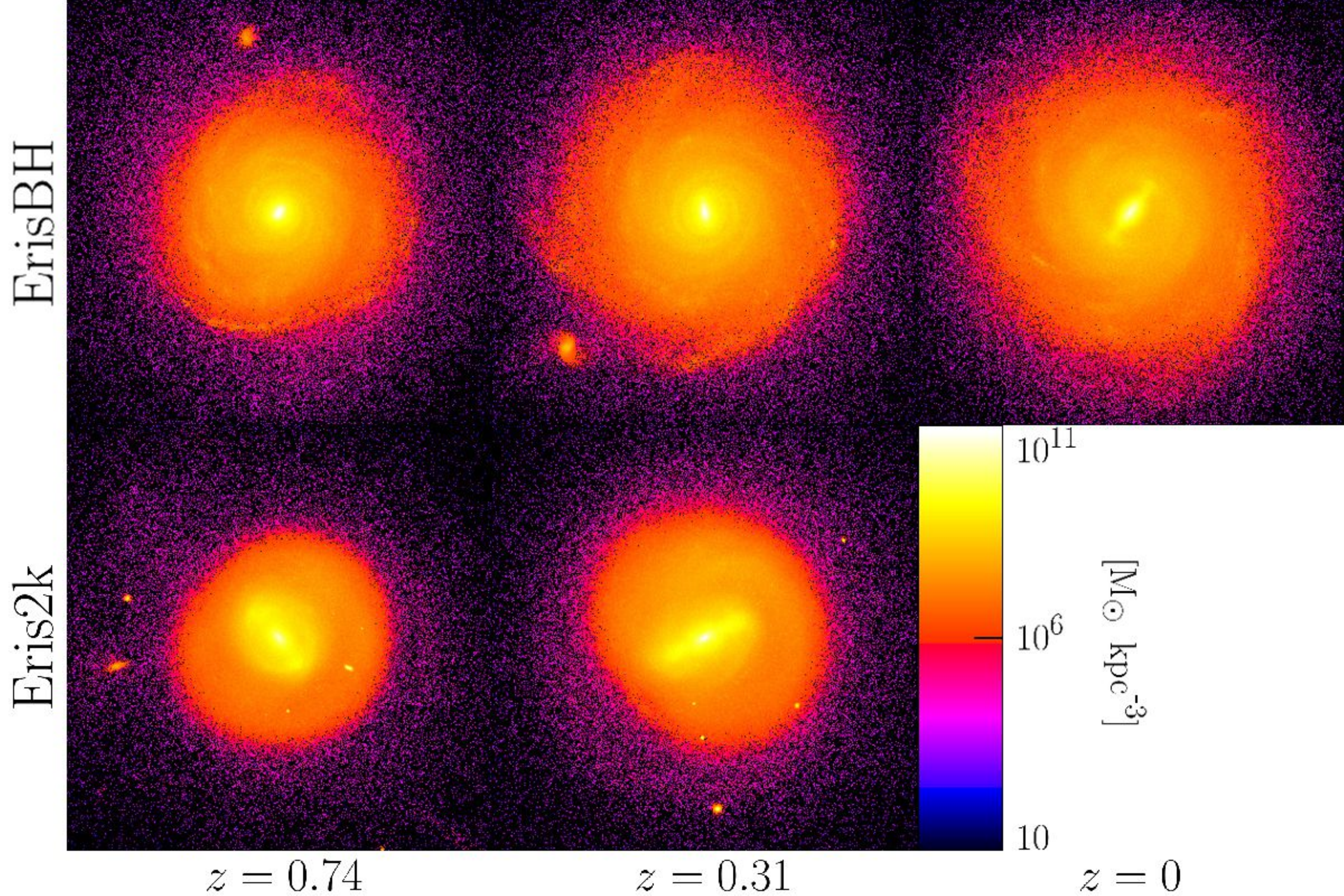}
\caption{Comparison between the stellar density maps of the main galaxies in ErisBH (first and third rows) and Eris2k (second and fourth rows) simulations. The boxes measure 40~kpc per side and have a fixed logarithmic colour scale, ranging from 10 to $10^{11}~\msun~{\rm kpc}^{-3}$. In the first block on top (six panels), we show, from left to right, the systems at $z=3$, $2$, and $1.14$, which is when the bar starts growing in Eris2k (see Section~\ref{sec:formation}). In the lower block (five panels), the redshift are $z=0.74$ (the ErisBH bar formation time), $0.31$ (the last snapshot of Eris2k), and $0$, which is present only for the ErisBH simulation, since the Eris2k run has been halted at $z \sim 0.3$.
In order to better appreciate the growing bars, we provide in Figures~\ref{fig:proj_form} and \ref{fig:proj}, the three projections of the disc central $12$~kpc at the time of bar formation and at the end of the two simulations, respectively.}
\label{fig:maps}
\end{figure*}

The Kron radius grows during the evolution of the discs from about $1$ to about $4$~kpc in both the runs and, at each redshift, the difference between the two $R_{\rm K}$ remains within $\sim$10 per cent of each other.
On the other hand, the stellar mass of the two galaxies (both within $2$ and $20$~kpc) can differ by almost a factor of two.
This hints to a far smaller effect of the different stellar-physics prescriptions over the disc extent,  compared to other fundamental properties, such as the stellar mass.
 A comparison between the two runs at different times is presented in Figure~\ref{fig:maps}.
It is worth emphasising here a fundamental difference between the studies of idealised isolated disc galaxies and cosmological studies: whereas in the former case the galaxies evolve for up to $\sim$10~Gyr \citep[e.g.][]{Athanassoula_et_al_2013} as if they formed from a monolithic collapse at the dawn of times, here each galaxy undergoes a significant evolution in mass and size, even during the last evolutionary stages when a growing bar is present. This gives us the unique opportunity to link the evolution of sub-structures, in particular the bars, to the cosmological growth history of the galaxies. 

\section{Feedback effects on sub-structures: the different lives of bars}
\label{sec:local_scales}

\begin{figure}
\vspace{1pt}
\includegraphics[width=\columnwidth]{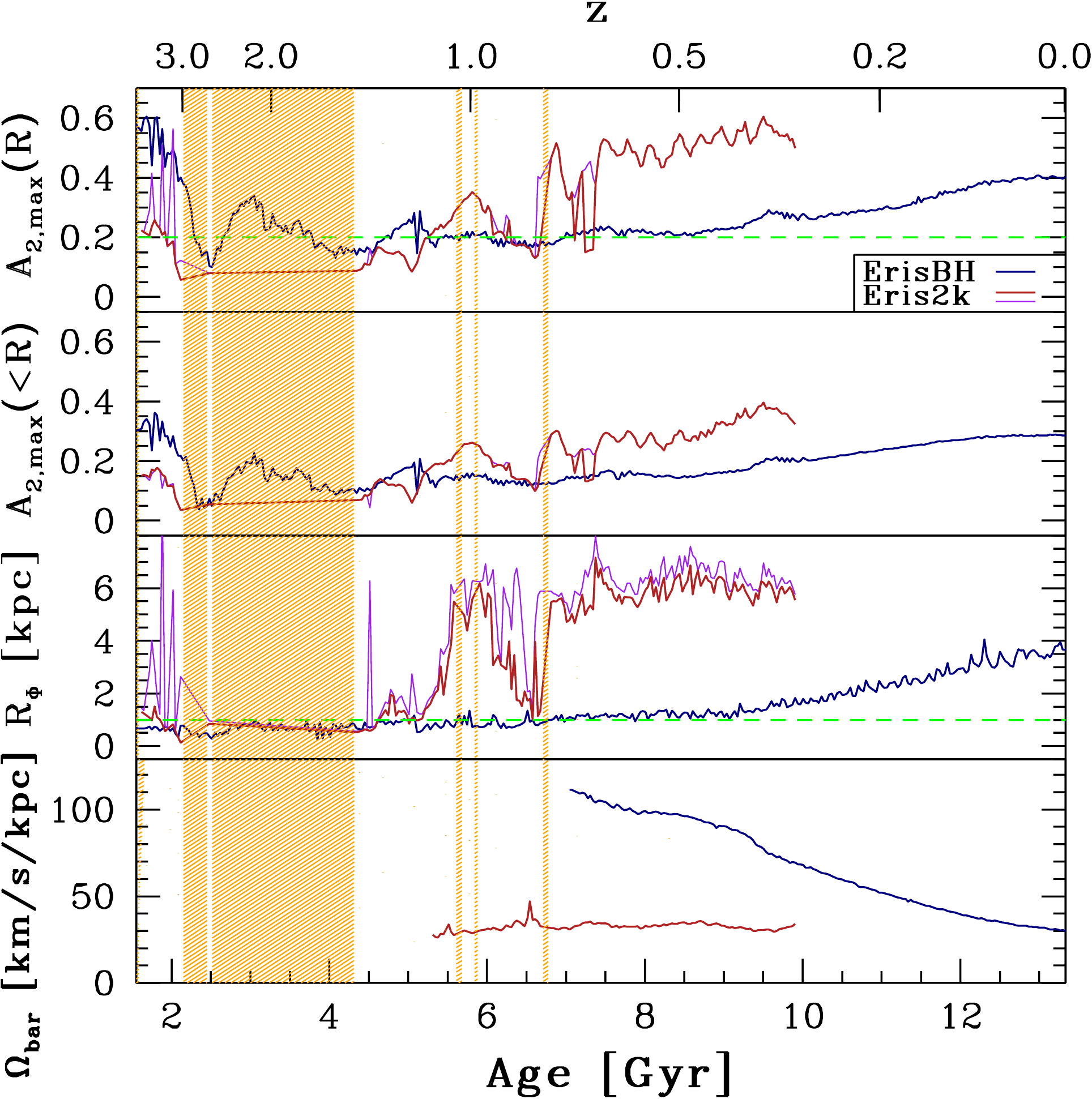}
\caption{Redshift evolution of the bar properties. From top to bottom: maximum strength of the two-fold deviation from axisymmetry measured in thin radial annuli (first panel) and averaged within any given radius $R$ (second panel); radial extent of such deviation (third panel), and its corresponding angular frequency (fourth panel). The shaded areas mark the redshift ranges where it was impossible to retrieve the bar properties for the galaxy in Eris2k. The dashed horizontal lines in the first and third panel mark the thresholds [$A_{2, \rm{max}}(R) = 0.2$ and $R_{\Phi} = 1$~kpc] assumed for the strong-bar classification. 
The purple lines show the Eris2k values for the overdensities marked as bars when $\left| \Phi(R_{\rm peak})-\Phi(R)\right|<\Delta \Phi =\arcsin(0.3)$, as opposed to $\Delta \Phi = \arcsin(0.15)$ in all other cases.}
\label{fig:ev}
\end{figure}

We performed a Fourier decomposition of the face-on view of the stellar surface density $\Sigma_*$ in order to quantify the formation epochs, strengths, lengths, and angular speeds of the bars forming in the two simulated galaxies. More specifically, we computed the local strength of any bar-like deviations from axisymmetry through the quadrupole-to-monopole ratio of the Fourier development:

\begin{equation}\label{eq:A2}
A_2(R) \equiv \frac{\left|\sum_{j}{m_{j}e^{2i\theta_{j}}}\right|}{\sum_{j}m_{j}},
\end{equation}

\noindent where $\theta_{j}$ is the azimuthal angle of the $j$-th particle of mass $m_{j}$ in the disc plane and the summation is carried over all the stellar particles enclosed in an annulus of width $\sim$10~pc, height 2~kpc, and centred at the radius $R$\footnote{We tested that the number of bins has a minimal influence over
the estimation of the bar parameters, as long as the bin-size is large enough to prevent strong statistical fluctuations.
The choice of the bin size has been made for consistency with our previous works \citep[see][]{Zana_et_al_2018a, Zana_et_al_2018b}.}.
Analogously to what has been done in \cite{Zana_et_al_2018a}, we also provide an averaged bar strength parameter $A_2(<R)$, evaluated through Equation~\eqref{eq:A2}, but including all the particles enclosed within the radius $R$. The maxima of the two profiles -- \mbox{$A_{\rm 2,max}(R) \equiv \max[A_2(R)]$} and \mbox{$A_{\rm 2,max}(<R) \equiv \max[A_2(<R)]$} -- are used as estimates of the bar strength, and their evolution with time is shown in the two uppermost panels of Figure~\ref{fig:ev}.

We point out that the sub-structures resulted in these simulations should be taken as prototypes of bars that can be formed in a cosmological context. We do not intend to perform a straight comparison between the two bars at a given time or radius.

\begin{figure*}
\includegraphics[width=0.94\textwidth]{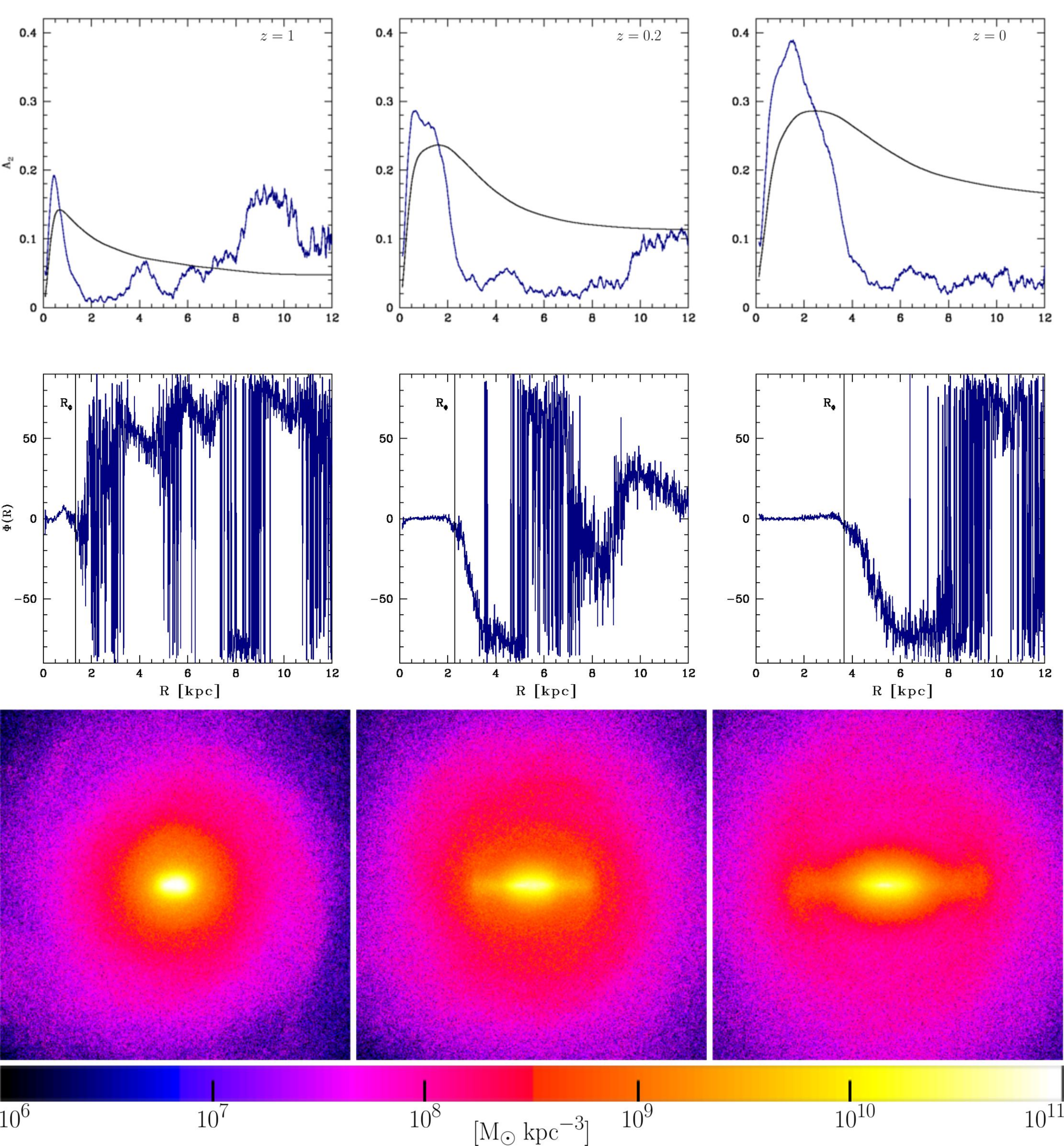}
\caption{Three snapshot of ErisBH, with increasing time from left to right. We show in the uppermost row the evolution of the $A_2(R)$ profile (blue lines) along with its cumulative counterpart $A_2(< R)$ (black lines) and in the middle row the profiles of the phase $\Phi(R)$ (blue lines) as they are computed through Equation~\eqref{eq:phase}.
The black vertical lines mark the positions of the bar length estimator $R_{\Phi}$.
As the redshift approaches $z=0$, the peaks of both the $A_2$ functions become higher (the bar gains in strength) and move farther out (the bar gains in length). At the same time, the plateau in $\Phi(R)$ becomes longer.
The growth of the structure is echoed in the corresponding stellar density maps (lowermost row). 
The logarithmic colour scale ranges from $10^{6}$ (black) to $10^{11}~\msun~{\rm kpc}^{-3}$ (white) and the side of each panel measures only $12$~kpc, in order to focus on the bar region.}
\label{fig:a2prof}
\end{figure*}

The sizes of the growing bars and their angular speeds are obtained analysing the radial profile of the angular phase of any two-fold asymmetry:

\begin{equation}\label{eq:phase}
\Phi(R) \equiv \frac{1}{2} \arctan \left[ \frac{\sum_{j}{m_{j}\sin(2\theta_{j})}}{\sum_{j}{m_{j}\cos(2\theta_{j})}} \right],
\end{equation}

\noindent where the sum is performed over the particles within narrow radial annuli as was done for $A_2(R)$. Wherever a bar-like structure (i.e. a straight $m=2$ mode) is present, the profile of $\Phi(R)$ shows a plateau (see, for instance, the middle row of Figure~\ref{fig:a2prof}). The length of such asymmetry $R_\Phi$, whose evolution is shown in the third panel of Figure~\ref{fig:ev}, has been estimated by checking for the extent of such plateau. Operatively, $R_\Phi$ is defined as the radius at which $\Phi(R)$ deviates from $\Phi(R_{\rm peak})$ by more than $\arcsin(0.15)$, where $R_{\rm peak}$ is the radius corresponding to $A_{\rm 2,max}(<R)$.\footnote{We note that a first prescription of such a kind has been discussed in \cite{Athanassoula_Misiriotis_2002}, where they checked for deviations from the cumulative value of $\Phi$ integrated out to the outermost regions of the galaxy. This is an optimal prescription for the study of idealised isolated galaxies, where no satellites and other cosmological structures are present. The value of the phase at the peak of $A_2(<R)$ has been originally proposed by \cite{Zana_et_al_2018b} as reference exactly to avoid the contamination from (minor) mergers, flybys, and other substructures within the disc.}

Figure~\ref{fig:a2prof} exemplifies how the Fourier decomposition method allows to follow a growing non-axisymmetry showing, for three ErisBH snapshots of decreasing redshift (from left to right), both the $A_2(R)$ (blue lines) and the $A_2(<R)$ (black lines) profiles (uppermost row), the phase profile $\Phi(R)$ (middle row) and the corresponding stellar density maps (lowermost row).
The longer and clearer the barred overdensity becomes, the higher the peaks in both the profiles of $A_2$ are and their positions move toward larger radii.
A clear evolution is also visible in the progressive straightening of the phase, which, in turn, results in a gradual increase of the length $R_{\Phi}$ (black vertical lines in the middle row).

Whereas the procedure to identify the bar is applied without restrictions on the ErisBH run, the bar in the Eris2k simulation appears less defined and the related surface density profiles are, in general, more noisy. 
As a consequence, the associated analysis requires some additional steps, which are detailed in Appendix~\ref{sec:find}.

\noindent The sudden drops of the red lines of Figure~\ref{fig:ev}, like the one at $\sim$7~Gyr, are signs of the ``clumpiness'' of the Eris2k surface density map (see Figure~\ref{fig:maps}). Some of them are absent in the purple lines, which provide an alternative estimate for the quantities $A_{\rm 2,max}(R)$, $A_{\rm 2,max}(<R)$, and $R_\Phi$, using \mbox{$\left| \Phi(R_{\rm peak})-\Phi(R)\right|<\arcsin(0.3)$ instead of $\arcsin(0.15)$}.
Even if it is clear that the differences between the two estimates are minimal, we notice that the drops are just numerical and can be removed, e.g. by increasing the maximum variation allowed to $\Phi(R)$. Unfortunately, this comes at the price of losing accuracy in the determination of $R_{\Phi}$, and we decided to keep $\arcsin(0.15)$.

Whenever a bar is present, its angular speed $\Omega_{\rm bar}$ (lowest panel of Figure~\ref{fig:ev}) is computed using the values of $\Phi_{\rm bar}$ [obtained by averaging $\Phi(R)$ over the annuli that are part of the bar] between consecutive snapshots.

In order to avoid possible effects due to the sub-optimal resolution and/or misinterpreting transient deviations from axisymmetry (caused, e.g. by a self-gravitating object not originated from the disc), we impose very conservative requirements for the detection of a {\it strong} bar in the analysed runs.
We identify a strong bar when
\begin{enumerate}
    \item $A_{\rm 2,max}(R)>0.2$;
    \item $R_{\Phi}>1$~kpc (about 10 softening lengths for $z<9$).
 \end{enumerate}

\begin{figure*}
\includegraphics[width=0.89\textwidth]{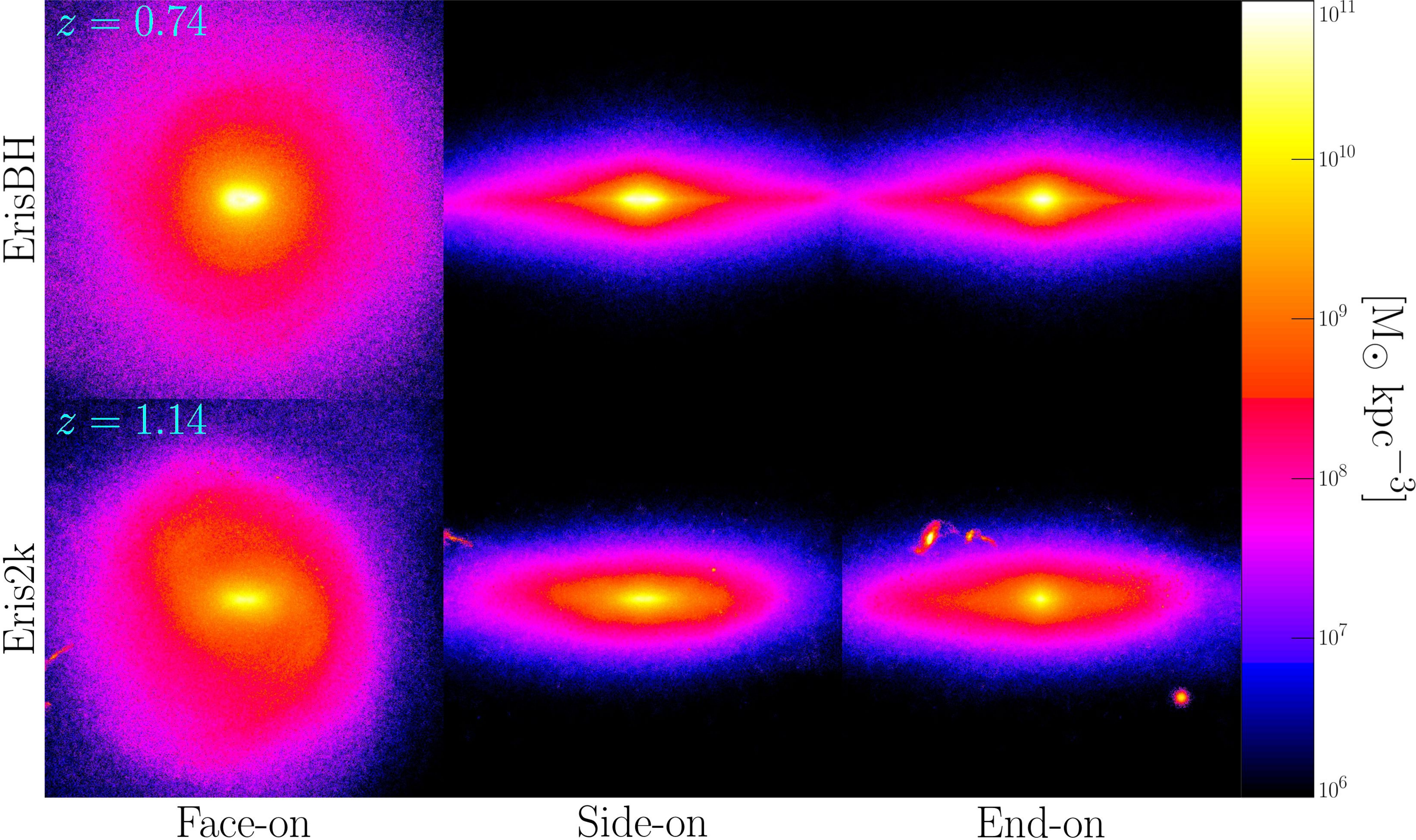}
\caption{From left to right: face-on, side-on, and end-on projections of the central region ($12$~kpc per side) for both the simulations ErisBH (top row) and Eris2k (bottom row) at the time of formation of their bar, according to the criteria we discuss in the text. The logarithmic colour scale ranges from $10^{6}$ (black) to $10^{11}~\msun~{\rm kpc}^{-3}$ (white).
The bars are barely discernible in these snapshots, but they soon increase their strength and length, following the evolution examined in Section~\ref{sec:local_scales}, to culminate in the fully grown structures visible in Figure~\ref{fig:proj}.}
\label{fig:proj_form}
\end{figure*}

These criteria result in a bar formation epoch of \mbox{$z \sim 1.14$} for Eris2k and \mbox{$z \sim 0.74$ }for ErisBH\footnote{A small (barely resolved) non-axisymmetric structure can be detected even at higher redshift, especially in the case of ErisBH; see also \citet{Guedes_et_al_2013} for a similar result in the case of the Eris simulation.} (the conditions are also briefly met at $z\sim1.2$, but this is due to the last minor merger; see \citealt{Bonoli_et_al_2016}).
Figure~\ref{fig:proj_form} shows the stellar density maps of both the galaxies at the epoch of bar formation, according to our constraints. In the central regions of both galaxies the origins of two elongated overdensities are recognizable, especially in face-on viewed maps (left-hand panels).

The evolutions in strength, length, and speed of the two forming bars are remarkably different. Eris2k has a significantly faster growth, with the bar length reaching a close to constant value in less than 1~Gyr. The sudden drop in the strength and length of the bar at $z \sim 0.95$--0.8 has been studied in \cite{Zana_et_al_2018b}, and is caused by the temporary shuffling of the orbits of the stars building the bar, due to the close passage (pericentric distance of $\sim$ 6.5~kpc) of a small satellite (of mass $\sim 1.1 \times 10^8 \msun$). As discussed in \cite{Zana_et_al_2018b}, the satellite passage does not modify substantially the primary potential profile nor, as a consequence, $\Omega_{\rm bar}$, and the bar regains its pre-interaction properties within $\sim$ 1~Gyr. Only close to the end of the run (age $>9$~Gyr in Figure~\ref{fig:ev}), the Eris2k bar starts to weaken and shorten, due to a bar-driven strong inflow of gas as in the bar-suicide scenario (see Section~\ref{sec:local_scales2}).

The ErisBH bar, on the other hand, shows a later start and a slower evolution, with strength and length gradually increasing till $z \sim 0$, when the growth in $A_2(R)$ and $A_2(<R)$ slows down due to the vertical instabilities and the consequent buckling of the bar (see Section~\ref{sec:local_scales2}).\footnote{The difference between $A_2(R)$ and $A_2(<R)$ is hardly noticeable in Eris2k, whereas it is clearer in the ErisBH run. This is due to the different mass concentrations in the stellar-dominated galactic nuclei: the stellar mass within $400$~pc is up to three times larger in the ErisBH case (see Section~\ref{sec:local_scales2}), affecting the monopole term of the Fourier development.} $\Omega_{\rm bar}$ shows a slow and approximately constant decrease, as commonly found in isolated systems \citep[see, e.g.][]{Athanassoula_2003}.\footnote{Here the angular momentum is redistributed from the inner zones toward the outskirts and this process drives the slowdown of the bar pattern speed and its increase in length.}

\begin{figure*}
\includegraphics[width=0.89\textwidth]{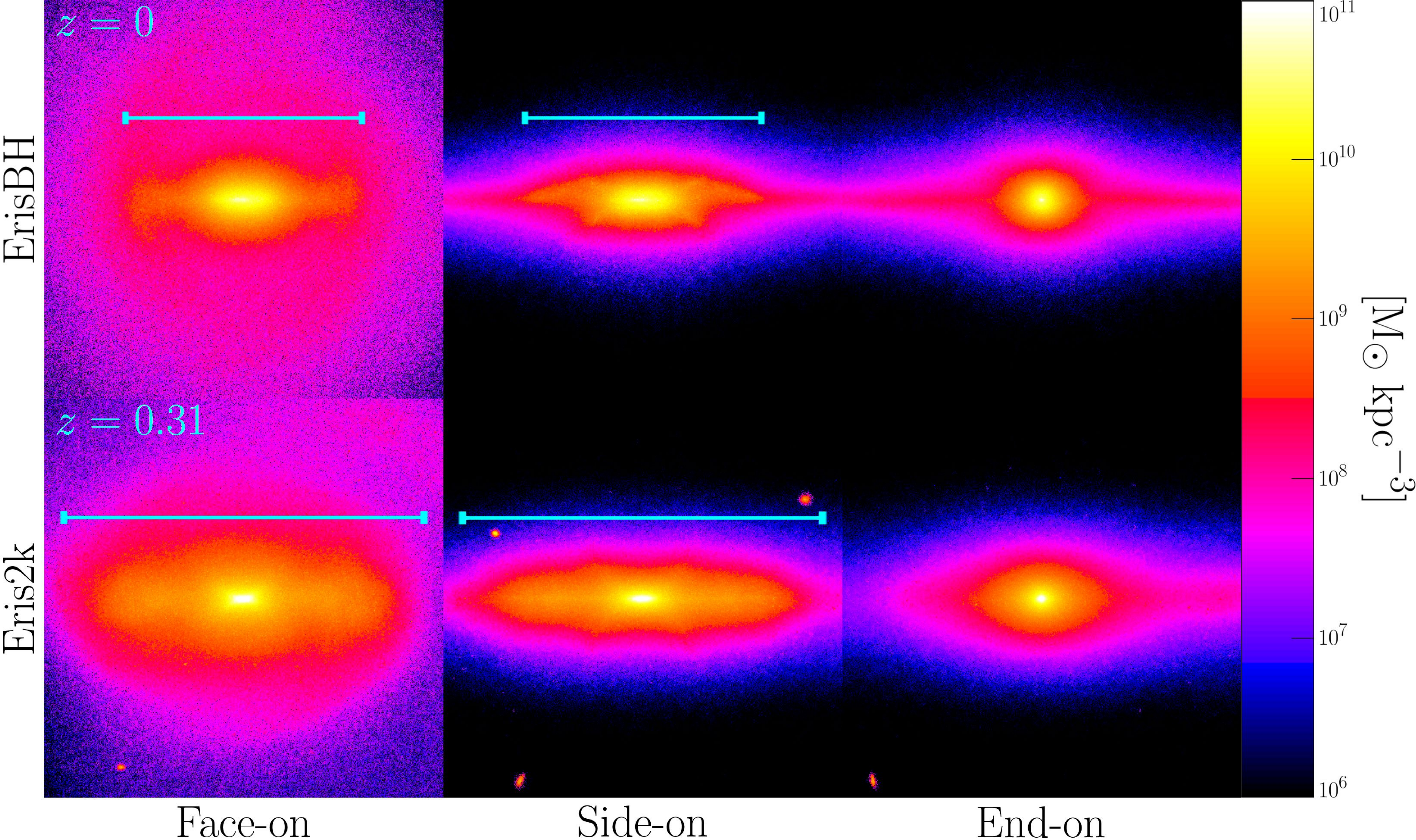}
\caption{Same as Figure~\ref{fig:proj_form}, but at the final snapshots of the two simulations.
The colour scale and size are kept identical to those of Figure~\ref{fig:proj_form} in order to facilitate the comparison.
At the represented times the bar (semi-)length $R_{\Phi}$ are $3.7$~kpc in ErisBH and $5.5$~kpc in Eris2k. The cyan rods correspond to twice $R_{\Phi}$.
The fully evolved structures almost fill the related panels, whereas the characteristic X-shaped pseudobulge is noticeable in the centre of both the galaxies as an index of the bar development stage. This is larger in the case of ErisBH, which also shows the signs of a progressive vertical asymmetry, completely missing in Eris2k (see Section~\ref{sec:local_scales2}, for a discussion of this process).}
\label{fig:proj}
\end{figure*}

\section{Feedback effects on sub-structures: the different deaths of bars}\label{sec:local_scales2}

\begin{figure}
\vspace{1pt}
\includegraphics[width=\columnwidth]{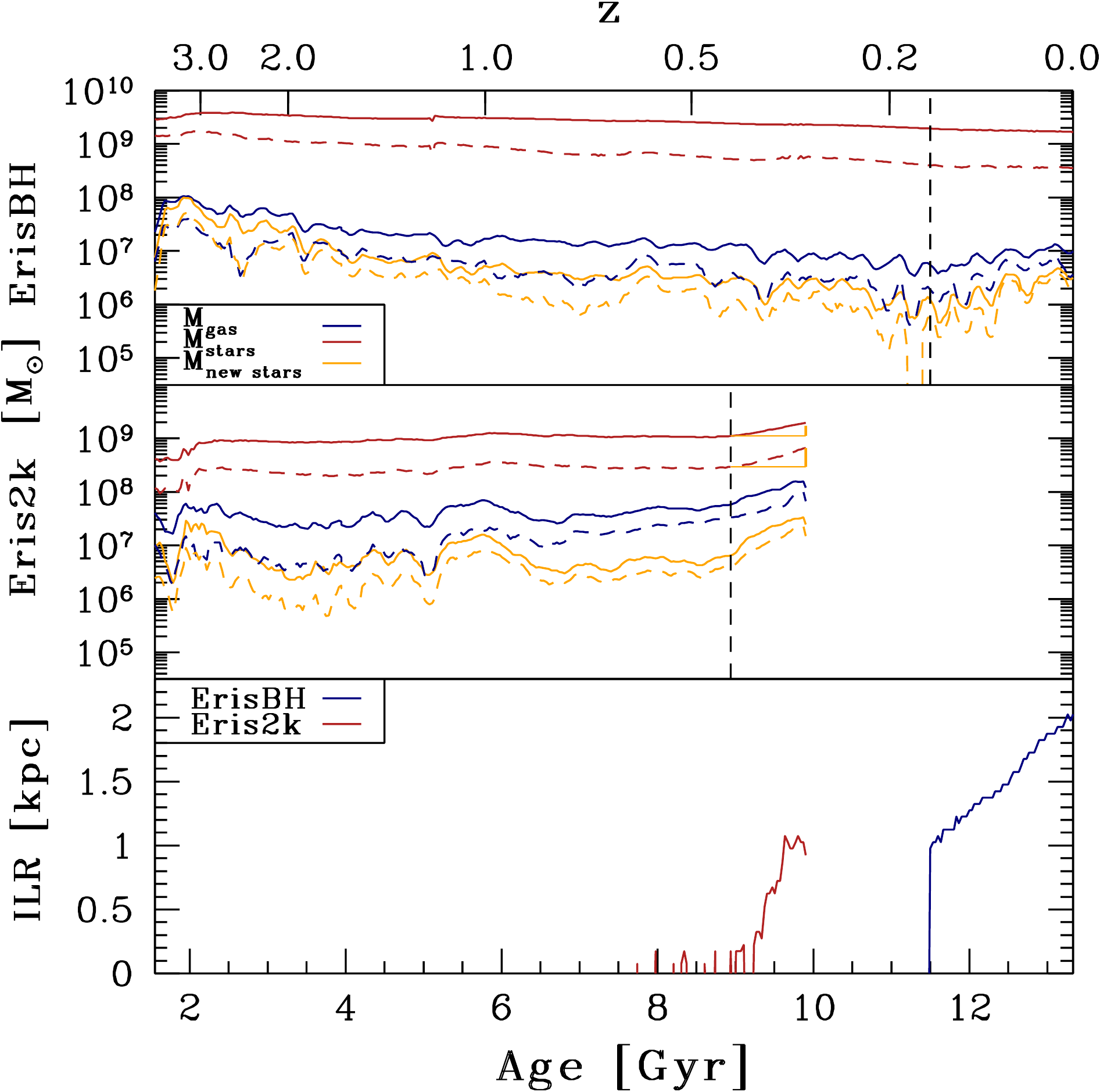}
\caption{Upper two panels: comparison between the central stellar (red lines) and gas (blue lines) masses in ErisBH (first panel) and Eris2k (second panel). Yellow lines refer to newly formed stars only (i.e. stars formed within 33~Myr, in between two snapshots). Solid and dashed lines refer to the mass within 0.4 and 0.2~kpc, respectively. 
The time when the ILR starts to arise is marked with a vertical black dashed line (see the discussion in the main text). 
The mass in stars formed after $z=z_{\rm{ILR}}$ (yellow vertical segment in the figure) is a significant fraction of the final stellar mass in the nucleus at $z \sim 0.3$.
Lower panel: evolution of the ILR radii for ErisBH (blue line) and Eris2k (red line).}
\label{fig:cmc}
\end{figure}

The weakening and shortening of the Eris2k bar at $z \lsim 0.4$ is related to the strong bar-driven gas inflow within the central 400~pc observable in the middle panel of Figure~\ref{fig:cmc}. It must be noted that, at higher redshift, the mass of the galactic nucleus in Eris2k remains significantly lower than that in ErisBH, due to the same higher impact of SN feedback that delays the overall growth of the galaxy with respect to its ErisBH counterpart. Only when the galaxy potential well is deep enough and the bar is already fully developed, the gas flowing toward the central regions of the galaxy can efficiently form a central stellar knot. This gas transfer leads to the so-called ``bar-suicide'' process \citep[e.g.][]{Pfenninger_Norman_1990, Norman_et_al_1996, Berentzen_et_al_1998, Shen_Sellwood_2004, Athanassoula_2005, Debattista_et_al_2006}, which is when the fast differential precession at different radii unravels the stars in the inner regions of the bar, decreasing its strength. 

\begin{figure}
\vspace{-6pt}
\includegraphics[width=\columnwidth]{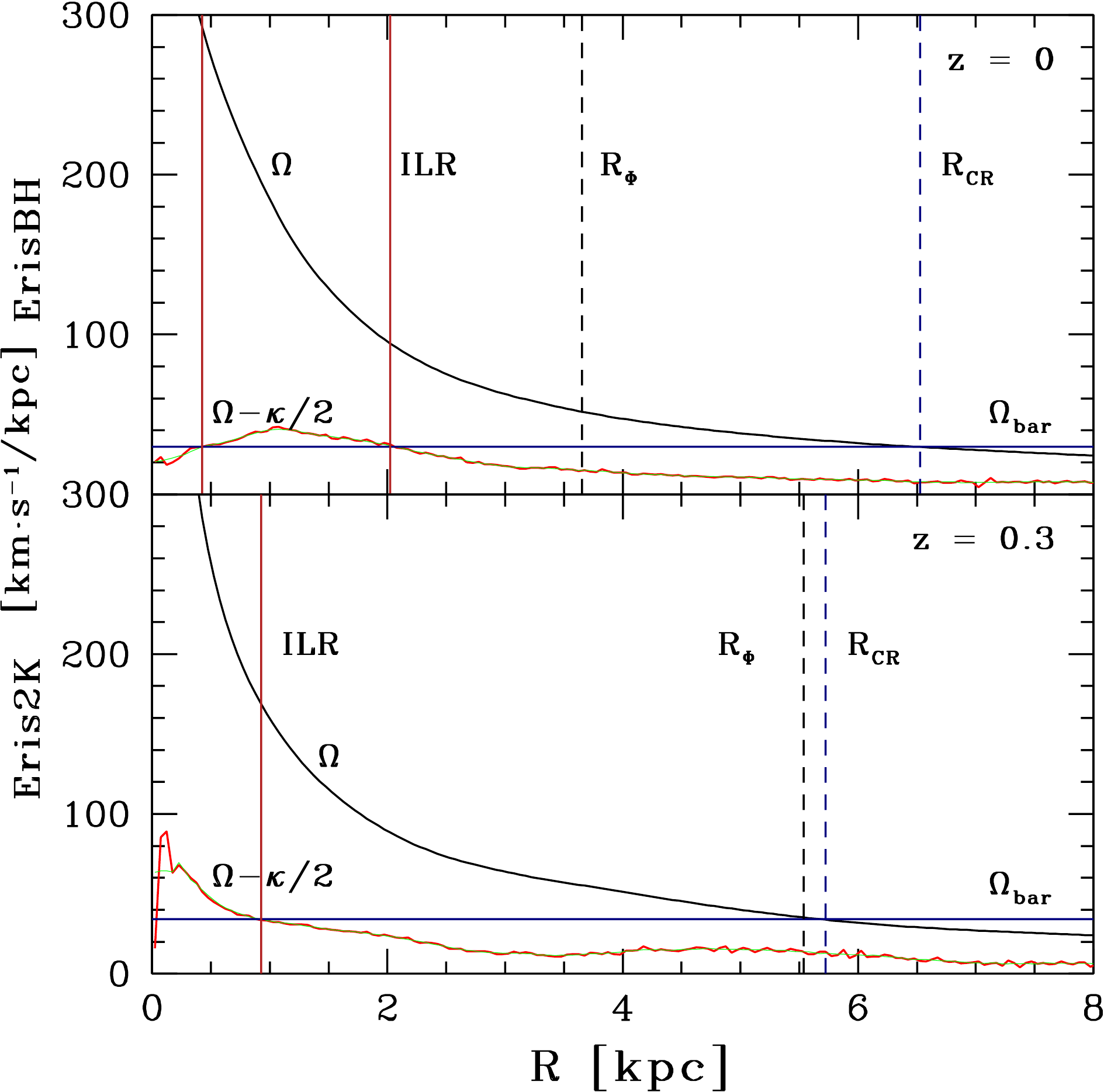}
\caption{Frequency maps of the main galaxies in ErisBH (upper panel) and Eris2k (lower panel) at the end of the two runs. 
The black and red curves refer to the angular velocity $\Omega(R)$ and to the precession frequency $\Omega(R)-\kappa(R)/2$, respectively. 
The horizontal blue lines refer to the bar angular velocities $\Omega_{\rm{bar}}$. 
The green lines show a smoothing of the precession frequency curves that have been used to find the intersections with $\Omega_{\rm{bar}}$. 
The black dashed lines refer to the bar extent ($R_{\Phi}$), whereas the other vertical lines highlight the positions of the corotation radii $R_{\rm{CR}}$ (that mark the radial position where the bar speed is exactly equal to the galactic rotation curve, blue dashed lines) and of the ILR (solid red lines).
}
\label{fig:prec}
\end{figure}

The effect of such a dense stellar nucleus is clearly observable in the radial profile of the precession frequency $\Omega(R)-\kappa(R)/2$ of a test particle on (little-)eccentric orbits in the disc plane \citep[see, e.g.][]{Binney_Tremaine_2008}, where $\Omega(R)$ is the circular angular frequency and $\kappa(R)$ is the frequency of small radial oscillations, here computed in the epicyclic approximation. $\Omega(R)-\kappa(R)/2$ is strongly sensitive to any central mass concentration (CMC) and has been already used to put constraints on the central massive BH mass of a disc galaxy, even when the BH influence radius was poorly resolved \citep{Combes_et_al_2014}.
The profiles of $\Omega(R)$ and $\Omega(R)-\kappa(R)/2$ are shown in Figure~\ref{fig:prec} for the last snapshot of the two runs. Eris2k shows a central cusp in $\Omega(R)-\kappa(R)/2$, that drops only at radii comparable to the gravitational force resolution. Such peak is not present before the above-mentioned gas inflow and, as a consequence, no inner Lindblad resonance [ILR, defined by the intersection between $\Omega(R)-\kappa(R)/2$ and $\Omega_{\rm bar}$] was found until $z=0.43$, as shown in the bottom panel of Figure~\ref{fig:cmc}.\footnote{The combined mass of the newly formed stars (of about $3.7 \times 10^{7}~\msun$ within $0.2$~kpc and $6.2 \times 10^{7}~\msun$ within $0.4$~kpc) after $z=0.43$ is almost equal to the variation of the total stellar mass in the same regions and in the same period. It follows that the gaseous inflow is fully responsible for the increase in the CMC.}

The precession frequency profile is considerably different in the ErisBH case, where the increase of $\Omega(R)-\kappa(R)/2$ toward the centre starts earlier in time and at larger radii, due to the large mass concentration already present at high redshift, but does not keep growing to the smallest resolved scales, probably due to the BH feedback implemented in this run.\footnote{The influence radius of the BH in ErisBH is far from being resolved. If it were resolved, $\Omega(R)-\kappa(R)/2$ would show a central divergence at pc scales and an ILR would always be present at such scales.} As a consequence, the ILR in ErisBH appears only when the bar has slowed down enough to intersect the $\Omega(R)-\kappa(R)/2$ profile, as shown in the bottom panel of Figure~\ref{fig:cmc}.

The slowing down of the bar growth in ErisBH is not associated to any gas inflow, as demonstrated in Figures~\ref{fig:cmc} (the stellar mass within $400$~pc shows a decrease at low redshift) and \ref{fig:prec}, but it is due to the ``buckling'' of the central regions of the bar, when the radial motions get partially converted into vertical motions above the disc plane, breaking the symmetry of the stellar distribution with respect to the disc plane and weakening the bar (e.g. \citealt{Combes_Sanders_1981, Raha_et_al_1991, Martinez-Valpuesta_Shlosman_2004, Debattista_et_al_2004, Debattista_et_al_2006}; or \citealt{Lokas_et_al_2019}, for a different interpretation).

\begin{figure}
\vspace{-7pt}
\includegraphics[width=\columnwidth]{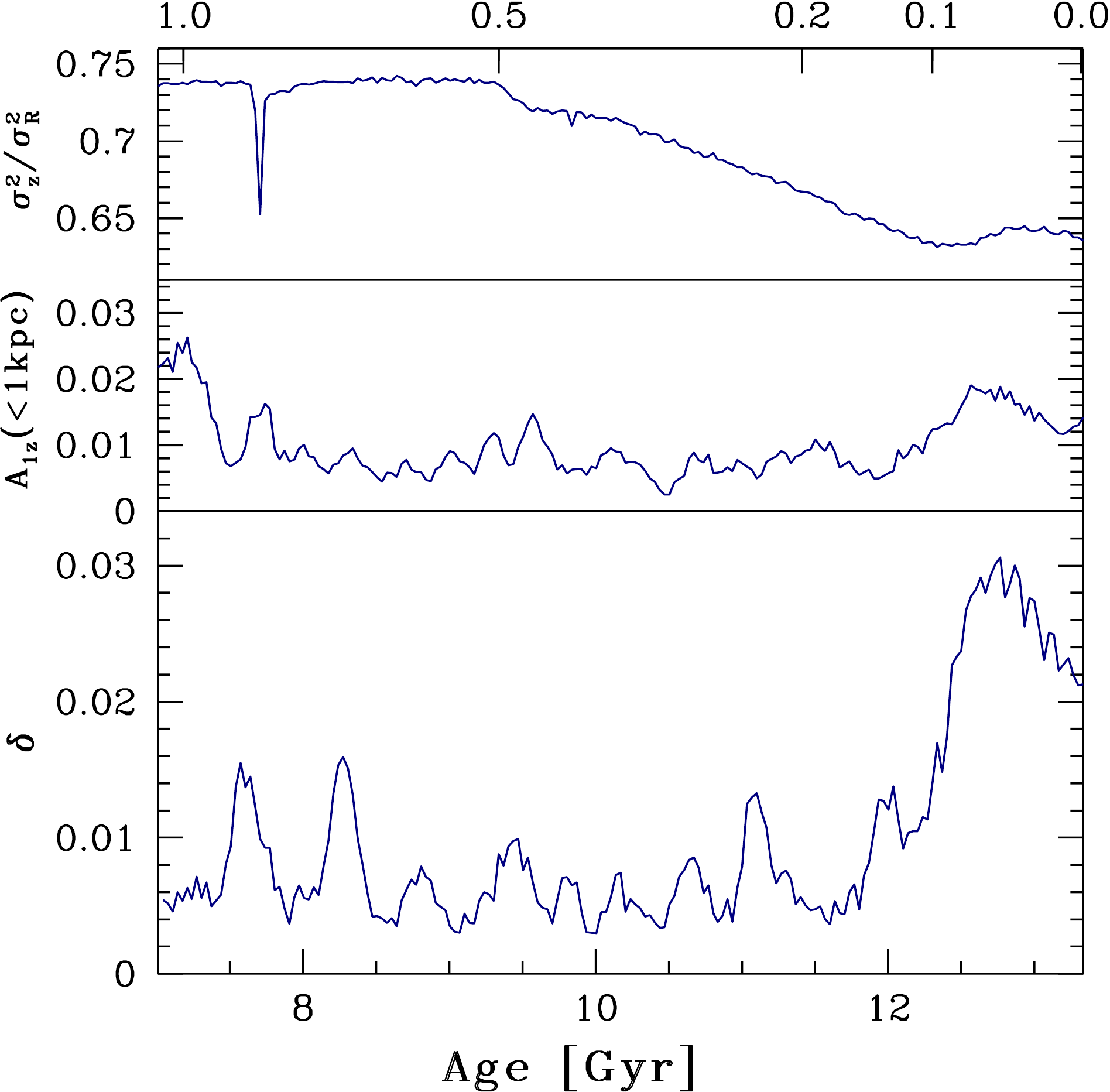}
\caption{Evolution of the buckling-instability estimator $\sigma_z^2/\sigma_R^2$ (uppermost panel), of the $m=1$ Fourier-based buckling-strength parameter $A_{1,z}(<1~{\rm kpc})$, and of the newly defined buckling-strength parameter $\delta$, for the ErisBH run.}
\label{fig:buckling}
\end{figure}

As originally suggested by \citet{Raha_et_al_1991}, the buckling is expected to occur when the ratio between the vertical and the radial velocity dispersion $\sigma_z^2/\sigma_R^2$ decreases below a given stability threshold. We follow \cite{Martinez-Valpuesta_et_al_2006} by computing $\sigma_z^2/\sigma_R^2$ on the stars that are in the bar only, i.e. by selecting only particles within 2~kpc from the disc plane and within 2~kpc from the bar's major axis. Isolated numerical models set the buckling-unstable threshold to $\sigma_z^2/\sigma_R^2\approx 0.6$ \citep{Sotnikova_Rodionov_2005}. The evolution of $\sigma_z^2/\sigma_R^2$ is presented for ErisBH in the upper panel of Figure~\ref{fig:buckling}, showing a decreasing trend as a function of time down to the above-mentioned threshold at $z \sim 0.1$. \cite{Martinez-Valpuesta_et_al_2006} performed a Fourier decomposition of the side-on stellar surface density\footnote{Defined as the edge-on projection of the galaxy, with the line of sight perpendicular to the bar major axis, referred to as $x$ hereby.} and selected the $m=1$ (over the $m=0$) mode as an estimate of the degree of buckling, according to
\begin{equation}\label{eq:A1z}
A_{1,z}(x<x_{\rm max}) \equiv \frac{\left|\sum_{j}{m_{j}e^{i\phi_{j}}}\right|}{\sum_{j}m_{j}},
\end{equation}
where the sum is performed only over the bar stars with the same geometrical limits adopted for the computation of $\sigma_z^2/\sigma_R^2$. 
The time evolution of $A_{1,z}$, estimated within $x_{\rm max}<1$~kpc, i.e. $A_{1,z}(<1~\rm{kpc})$, is shown in the middle panel of Figure~\ref{fig:buckling}. A net increase in the buckling parameter is clearly observable as soon as the galaxy becomes buckling-unstable, due to the breaking of symmetry with respect to the disc plane. This bump corresponds to an increase in $\sigma_z$ (hence in the $\sigma_z^2/\sigma_R^2$ parameter), observable in the upper panel at late times. The richness of small sub-structures of both cosmological and internal origin produces the fluctuations present in the evolution of $A_{1,z}(<1~\rm kpc)$.

We therefore engineer a new quantitative estimate for the degree of buckling: at any value of $x$, we first compute the height above and below the disc plane within which 90 per cent of the stellar mass is included [dubbed $z^+(x)$ and $z^-(x)$, respectively], applying again the same geometrical boundaries used to compute $\sigma_z^2/\sigma_R^2$ and $A_{1,z}$. We then quantify the buckling asymmetry by computing the $x$-averaged relative difference of the $z^+(x)$ and $z^-(x)$ profiles:  
\begin{equation}\label{eq:delta}
\delta=2\frac{\int_{-8 \,{\rm kpc}}^{8 \,{\rm kpc}} (z^{+}(x)-|z^{-}(x)|) {\rm d} x}{\int_{-8 \,{\rm kpc}}^{8 \,{\rm kpc}} (z^{+}(x)+|z^{-}(x)|) {\rm d} x}.
\end{equation}

The evolution of $\delta$ is shown in the bottom panel of Figure~\ref{fig:buckling}, where a clear prominent peak is observed at $z\lsim 0.1$, in agreement with the other estimators. The buckled part of the disc does evolve close to the end of the run into a boxy-peanut bulge, decreasing the asymmetry as observable both in $A_{1,z}(<1~{\rm kpc})$ and $\delta$, as already detailed for the ErisBH case in \cite{Spinoso_et_al_2017}.

\section{Dynamical interpretation of the different bar properties}
\label{sec:formation}

Two parameters are commonly \citep[see, e.g.][]{Ostriker_Peebles_1973, Combes_Sanders_1981, Sellwood_1981} considered to determine the susceptibility of a stellar disc to develop a bar. These are the $Q$ parameter \citep{Toomre_1964}, defined as
\begin{equation}
Q(R)\equiv\frac{\sigma_{\rm R}(R)\kappa(R)}{3.36G\Sigma_\star(R)},
\label{eq:toomre}
\end{equation}
where $\sigma_{\rm R}$ is the radial velocity dispersion and $G$ the gravitational constant, and the swing amplification parameter for an $m=2$ perturbation $X$ \citep{Goldreich_Tremaine_1978, Goldreich_Tremaine_1979}, defined as
\begin{equation}
X(R)\equiv \frac{k_{\rm crit}(R)R}{2} = \frac{\kappa^2(R)R}{4\pi G\Sigma_\star(R)},
\label{eq:swing}
\end{equation}
where $\lambda_{\rm crit}=2\pi/k_{\rm crit} = 4\pi^2 G\Sigma_\star \kappa^{-2}$ is the longest unstable wavelength\footnote{We are aware that $\lambda_{\rm crit}$ cannot measure directly the perturbation wavelength for a non-axisymmetric mode such as $m=2$. On the other end, it does provide an estimate of the scale size of the disc region that can become unstable to its own self-gravity.} \citep{Binney_Tremaine_2008}.
Although a precise value to determine the onset of a non-axisymmetric instability is not available, a common assumption (that we also use in what follows) is that spiral waves and bars can form for $1<Q\lesssim 2$ and $X\lesssim 3$ \citep[e.g.][]{Toomre_1964,Toomre_1981}.

\begin{figure}
\vspace{-13pt}
\includegraphics[width=\columnwidth]{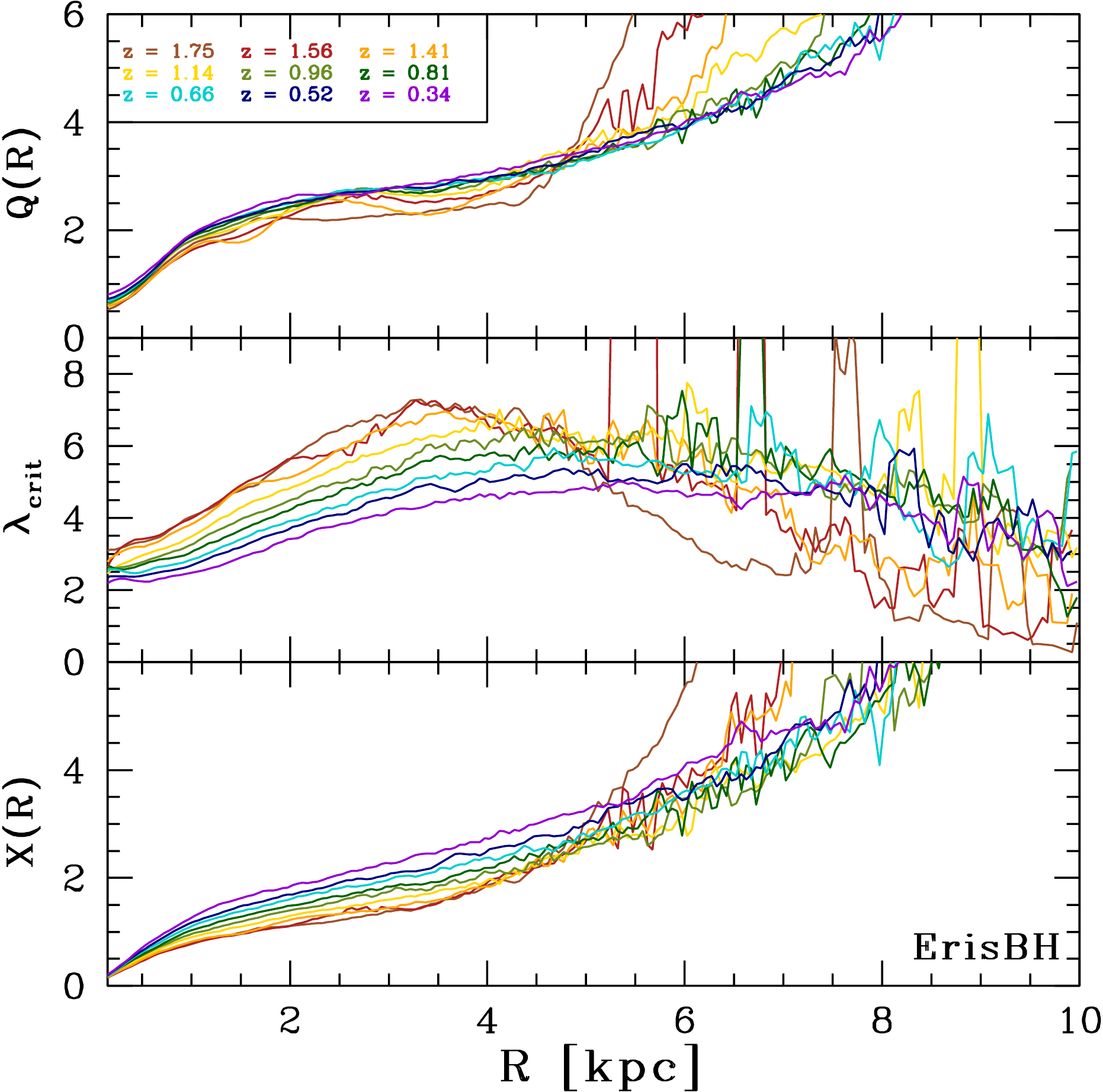}
\caption{Stability parameters for ErisBH (from top to bottom): Toomre parameter $Q$, critical wavelength $\lambda_{\rm crit}$, and swing amplification parameter $X$. The profiles are evaluated from the centre of the galaxy till $10$~kpc, over $1000$ equally spaced circular bins of height $8$~kpc. The different colours refer to different cosmic times, as it is reported in the upper panel with the corresponding colour code (from red to violet, the time is monotonically increasing).}
\label{fig:QXL_bh}
\end{figure}

\begin{figure}
\vspace{-13pt}
\includegraphics[width=\columnwidth]{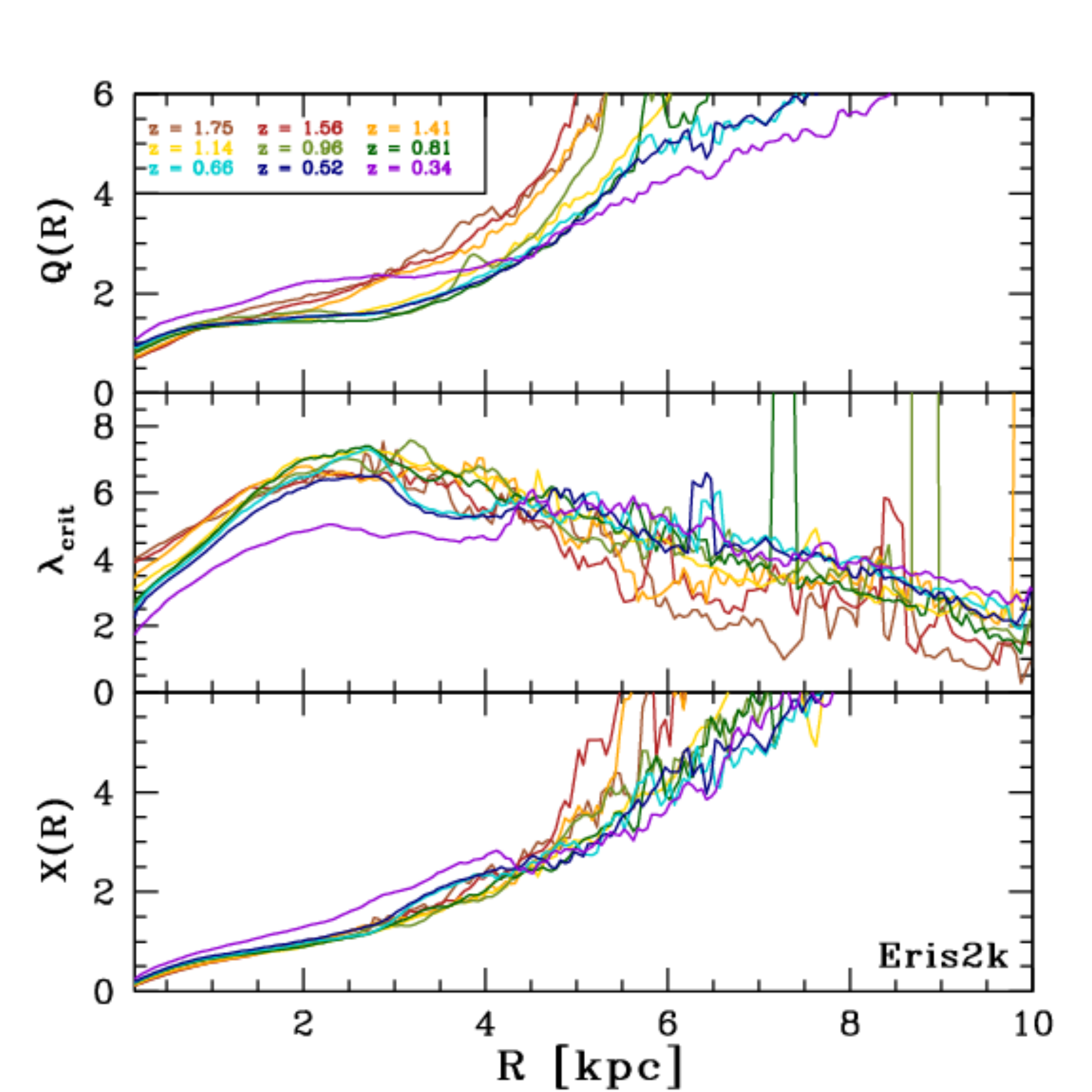}
\caption{Same as Figure~\ref{fig:QXL_bh}, but for Eris2k.}
\label{fig:QXL_2k}
\end{figure}

To assess the conditions for bar formation in Eris2K and ErisBH, we estimate $Q$, $X$, and $\lambda_{\rm crit}$ as a function of $R$ for different redshifts, from $z=1.75$, when the bar is not formed yet, down to $z=0.34$. The different quantities are reported in Figures~\ref{fig:QXL_bh} (ErisBH) and \ref{fig:QXL_2k} (Eris2K). 
We stress that both $Q$ and $X$ depend directly on the galactic potential, hence on the stellar distribution within the galaxy. Differences in these parameters are directly associated to the different (both in time and space) SF histories in the two runs, as discussed in Sections~\ref{sec:global_scales} and \ref{sec:local_scales}.

In ErisBH, $Q$ (top panels) does not vary significantly over the redshift range considered, showing a shallow radial profile with $Q\gsim 2.5$ between $R=1$ and $R=$ 4~kpc. In Eris2K, on the other hand, $Q$ monotonically decreases with redshift, settling around $Q \sim 1.5$ for $R<4$~kpc, and remains always lower than the ErisBH values.\footnote{The only exception in the evolutionary trend of Eris2K appears at $z\lesssim 0.4$, and is compatible with the bar weakening (see Section~\ref{sec:local_scales2}).}
We note that, in Eris2K, the flattening in the profile of $Q$ occurs when the bar forms, with $Q\gsim 3$ only above $R\sim 5$~kpc, corresponding roughly to the extension of the bar (see Figure~\ref{fig:ev}).
Assuming a critical value $Q_{\rm crit}\sim 2$ for global instabilities to develop, the observed trends confirm indeed that the bar in Eris2K, at formation time, is already larger and stronger than that in ErisBH, which is limited to the central kpc only.

$\lambda_{\rm crit}$ (middle panels) and $X$ (bottom panels), on the other hand, show a completely different evolution. As the galaxy evolves, $\lambda_{\rm crit}$ ($X$) in ErisBH exhibits a slow decrease (increase) that limits the maximum extension the bar could possibly reach during its evolution. In Eris2K, instead, both quantities remain (almost) constant, witnessing a negligible change of the bar properties. 

A comparison of the two figures shows that the (slightly) later formation time and the initial extension of the bar in ErisBH can be easily explained in terms of $Q$ and $X$, that are $\sim$ 1.5--2 times larger (for $R\lesssim 4$--5~kpc) than in Eris2K, making the disc in the latter case prone to stronger instabilities able to trigger the formation of a stronger and more extended bar.

During the evolution of the bar, the stability parameter analysis remains consistent with the picture in Figure~\ref{fig:ev}. In particular, the bar in ErisBH is initially small ($Q<2$ only within the central kpc), and grows up to $R_\Phi \sim 4$~kpc by $z=0$. The maximum extension in this case is limited by the decrease of $\lambda_{\rm crit}$, which peaks around 4~kpc at $z=0.34$. The bar in Eris2k, on the contrary, is fully developed from its start, because of $Q\lesssim 2$ up to $R\sim 4$~kpc and $\lambda_{\rm crit}$ peaking around 6--7~kpc, but does not evolve significantly with redshift. The only exception is at $z=0.34$, when the bar starts to dissolve, and the disc becomes stable also at smaller scales ($Q$ and $X$ start to increase and $\lambda_{\rm crit}$ drops to less than $4$~kpc).

It should be noted that the Toomre's stability Equation~\eqref{eq:toomre}, as well as the swing amplification Equation~\eqref{eq:swing} are derived in the Wentzel--Kramers--Brillouin (WKB) approximation,\footnote{In linear perturbation theory, the WKB approximation assumes that the long-range gravitational coupling is negligible with respect to local interactions, so that the response can be locally described \citep{Binney_Tremaine_2008}.} which is only reliable when $\lambda_{\rm crit}$ is short compared to the length of the system, or if $X>1$. Although our discussion is only aimed at the comparison of the evolving properties of the two simulations, it is clear from Figures~\ref{fig:QXL_bh} and \ref{fig:QXL_2k} that the WKB approximation is not satisfied for $R\lsim1$. The description of instability waves in these conditions would require a more advanced method in non-linear theory, which is beyond the scope of this work.

\section{Conclusions}
\label{sec:conclusions}

In this study, we presented a detailed comparison of the differences between two distinct cosmological zoom-in simulations starting with the same initial conditions. The different physical prescriptions on unresolved scales assumed in the two runs resulted in the formation of two disc barred galaxies whose bars show very different properties. 
A bar forms early ($z \sim 1.1$) in Eris2k, reaching a  size of $\gsim 6$~kpc within a very short initial growth phase of $\sim$1~Gyr. After such sudden growth, both the bar length and precession velocity remain approximately constant up to the final stages of  weakening due to a bar-triggered gas inflow as in the bar-suicide scenario. Large and fast bars, like that of Eris2k, are not uncommon in the Universe. Some examples are provided, for instance, by UGC 508, UGC 3013, or UGC 4422 \citep{Font_et_al_2017}.
On the other hand, the bar in ErisBH starts forming at slightly later times ($z \sim 0.7$) and keeps increasing its size (and decreasing its precession velocity) until the end of the run. The bar in ErisBH remains always smaller than its Eris2k counterpart, reaching a maximum size of $\sim$4~kpc at $z \sim 0$, when it stops growing due to a buckling event.

In \citet{Zana_et_al_2018a}, we demonstrated that the last minor merger in the ErisBH simulation does not provide the initial trigger necessary to the formation of the bar. However, the encounter can induce a delay in the formation time.
The difference between the bar formation epochs of ErisBH and Eris2k, being of the same order of magnitude of such a delay, could also be explained by a dynamical perturbation external to the galactic environment.
As a consequence, it is not clear whether the overall feedback produces any variation in the bar formation time, but it surely sensibly controls its structural properties by moulding the disc potential both on small and large scales.

The distinct bar evolutions and features are due to the different mass growths of the galaxy. The stronger effective feedback in Eris2k has the effect of initially pushing the gas out of the galaxy, more effectively reducing the SF, and results, in turn, in an initial stellar mass smaller than that in ErisBH.
The pushed-away gas flows back onto the main galaxy but, only when the disc is massive enough to prevent SN-driven massive gas ejections, Eris2k starts forming stars with high SF rate. 
The more efficient removal of low angular momentum gas at early times in Eris2k results in a lower CMC with respect to ErisBH. Such central stellar nuclei contribute in shaping the overall gravitational potential of the two galaxies, initially determining the size of the bar-unstable regions and, on the long run, the bar evolutions.

A recent study by \cite{Gavazzi_et_al_2015} has observationally proved a link between a knee mass ($M_{\rm knee}$) in the `specific SF rate versus stellar mass' plane for local star forming galaxies and the occurrence of strong bars at $z \sim 0$. They further speculated that the increased probability of forming a bar at high masses could explain the correlation they found between $M_{\rm knee}$ and $z$. The paucity of sufficiently high angular resolution images of stellar discs at cosmological distances prevents the statistical confirmation of such conjecture. Interestingly, the bars in ErisBH and Eris2k form when the galaxies become more massive than $M_{\rm knee}(z)$, as shown in Figure~\ref{fig:mass}. A larger statistical sample of high-resolution cosmological simulations of disc galaxies could populate the `bar-formation-mass versus $z$' plane, to theoretically probe the redshift-dependent mass threshold for bar formation proposed by \cite{Gavazzi_et_al_2015}.

It is noteworthy how well the feedback prescriptions can be connected to the different bar morphologies. Although the rigid constraints chosen to identify a strong bar (which are the subject of this investigation) are not fulfilled in the ErisBH run for $z > 0.7$, a small non-axisymmetric overdensity could anyway be observed, surely more pronounced with respect to the Eris2k equivalent galaxy (at the same redshift; see Figure~\ref{fig:ev}).
Despite the presence of this possible proto-bar (that can also be due to spurious numerical effects, as stressed in Section~\ref{sec:local_scales}), it is clear that the evolved bar at lower redshift remains close to the nuclear region, in contrast with the grand-scale bar of Eris2k.
These differences could be attributed to the distinct domains of the feedback mechanisms: whereas the effect of BH feedback is confined to the region $<1$~kpc, the strong effective stellar feedback in Eris2k acts on a global scale.

Even though we are limited by having analysed galactic bars from only two cosmological zoom-in simulations, it is enticing to speculate on the
general relationship between feedback and bars. From the results presented in this work, one can expect that, at low redshift ($z \lesssim 1$), bars are generally stronger and longer when the effective feedback is enhanced. Since strong/long bars are easier to observe, especially at $z \gtrsim 0.5$, the number of observed bars could give us some hints on the strength of effective feedback during the cosmological build-up of galaxies. We caution, however, that this comparison is possibly degenerate with other physical phenomena, such as the global merging history and environment.

In conclusion, this study clearly highlights a link between the structural properties of bars (and of the whole discs) and the sub-resolution physics implemented in simulations. 
This connection could be exploited to constrain and better tune the parameters of such implementations, possibly breaking the degeneracy with other free parameters such as, e.g. spatial and mass resolution or BH physics.
A large number of high-resolution cosmological simulations would be necessary for the purpose, also to isolate the influence of the external perturbations on to the processes of bar formation and evolution \citep[see][for an example on the Eris2k case]{Zana_et_al_2018b}.

\section*{Acknowledgements}
The Authors would like to thank the anonymous Reviewer for the helpful suggestions that improved the clarity of the manuscript.

\scalefont{0.94}
\setlength{\bibhang}{1.6em}
\setlength\labelwidth{0.0em}
\bibliographystyle{mnras}
\bibliography{bar_feedback}
\normalsize

\appendix

\section{Fitting method}
\label{sec:fit}

\begin{figure}
\includegraphics[width=\columnwidth]{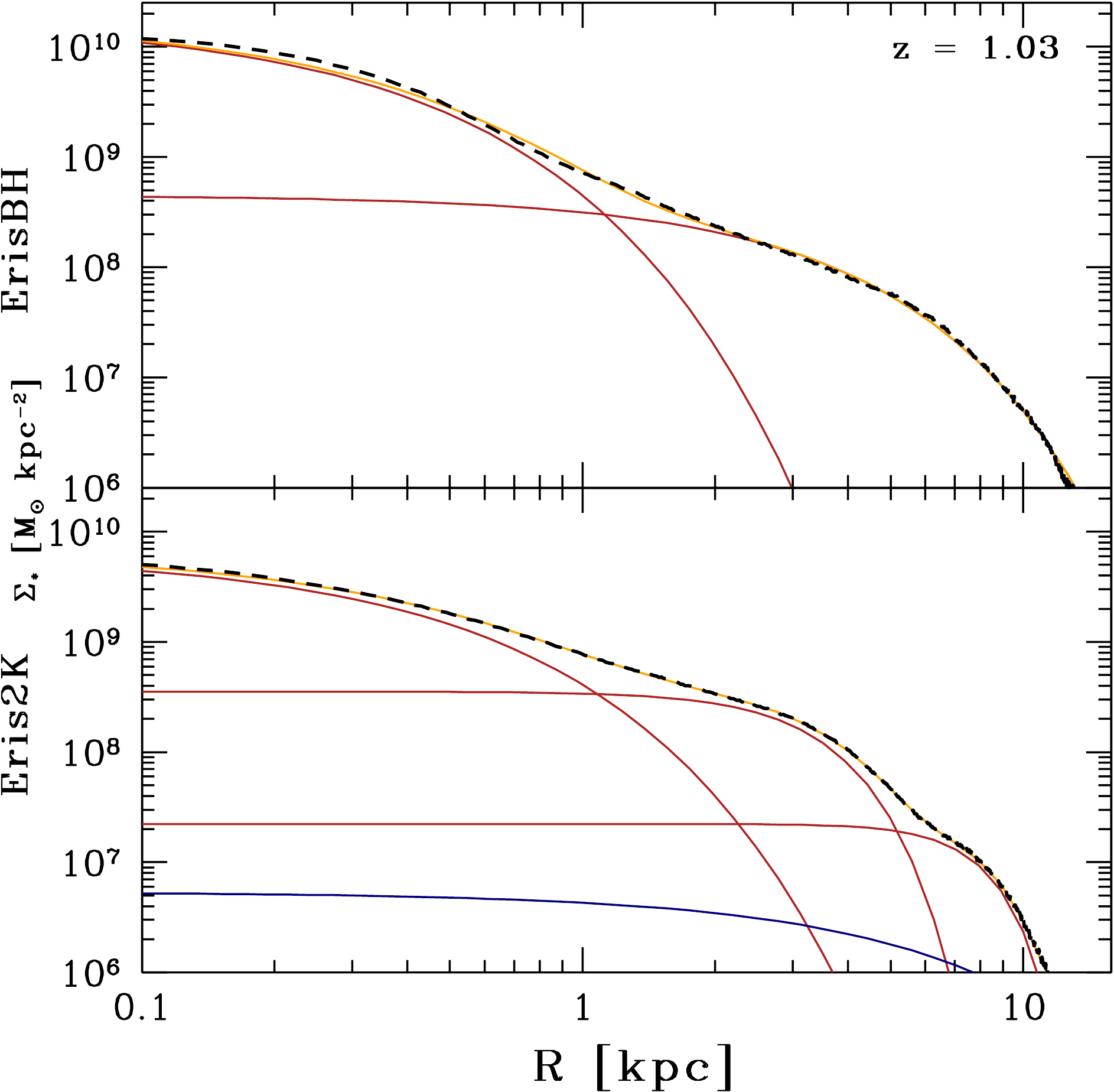}
\caption{Example of the fitting procedure applied onto the ErisBH (top panel) and Eris2k (bottom panel) main galaxies at $z=1.03$. The agreement between the surface density profiles $\Sigma_*$ (black dashed lines) and the best fits (yellow solid lines), where the relative error is lower than 15 (10) per cent for ErisBH (Eris2k), shows the goodness of the applied procedure. The red solid lines refer to the S\'ersic components, whereas the blue solid line in the bottom panel represents the exponential function we use to fit the stellar background in Eris2k.}
\label{fig:fit_2}
\end{figure}

In this section, we provide a brief explanation of the method we employ to produce the one-dimensional fits of the stellar distributions, exemplified in Figure~\ref{fig:fit} for the final snapshots of the two runs, and in Figure~\ref{fig:fit_2} at $z=1.03$.

First, a stellar surface density profile is extracted from each snapshot by dividing the disc into $20$~pc-wide concentric cylindrical bins, starting from the galactic centre.
The height of the bins measures 8~kpc in order to ensure that the entire galactic structure is included, minimizing at the same time the contamination by other systems.
The results are the black dashed lines of Figures~\ref{fig:fit} and \ref{fig:fit_2}.

Our decomposition procedure is based on a fitting algorithm from \citet{Press_et_al_1993} included in an iterative procedure discussed in the following.

For the case of ErisBH, the profiles are simpler with respect to Eris2k, and are typical of a disc galaxy with a small bulge/bar.
For this reason, we use a superposition of two \citet{Sersic_1963,Sersic_1968} functions in order to represent these two components (see the red lines in the top panel of Figure~\ref{fig:fit_2}).
The evolution of the galaxy is less disturbed by random encounters and its surface density shows a more predictable development.
Accordingly, the method is almost completely automatised and flawlessly allows to find a satisfactory fit for each snapshot. Once the first snapshot at $z=0$ is successfully fitted,\footnote{We start the decomposition from the last temporal snapshot, since it is, in general, easier to fit.} the resulting parameters are used as the initial guess for the previous (in time) contiguous snapshot and so on, towards higher redshift, when the trend of the density profiles becomes less and less trivial.

For Eris2k, given the higher complexity of the stellar mass distribution, a two-components fit is not sufficient, hence we select a four-components fit for the vast majority of the snapshots (see the bottom panel of Figure~\ref{fig:fit_2}) and a three-components fit for a few remaining snapshots at higher redshift.
Operatively, (i) we first fit the galaxy outskirt only (for $R>15$~kpc), using an exponential profile (blue solid line in the bottom panel of Figure~\ref{fig:fit_2}). Then (ii) we fit the inner regions with three (two for a few cases) S\'ersic profiles, keeping fixed the exponential background, previously extrapolated.
In this case as well, we initially focus on the last snapshot (at $z=0.31$), since its components are far more recognizable.
Hence, the procedure is mostly automatised (with minimal or no user intervention), by adopting the outcoming fitting parameters of each snapshot as the initial guess for the next snapshot to analyse.

The method is applied recursively for each profile, in order to achieve an even better final agreement.

\section{Finding the bar in Eris2k}
\label{sec:find}

\begin{figure}
\includegraphics[width=\columnwidth]{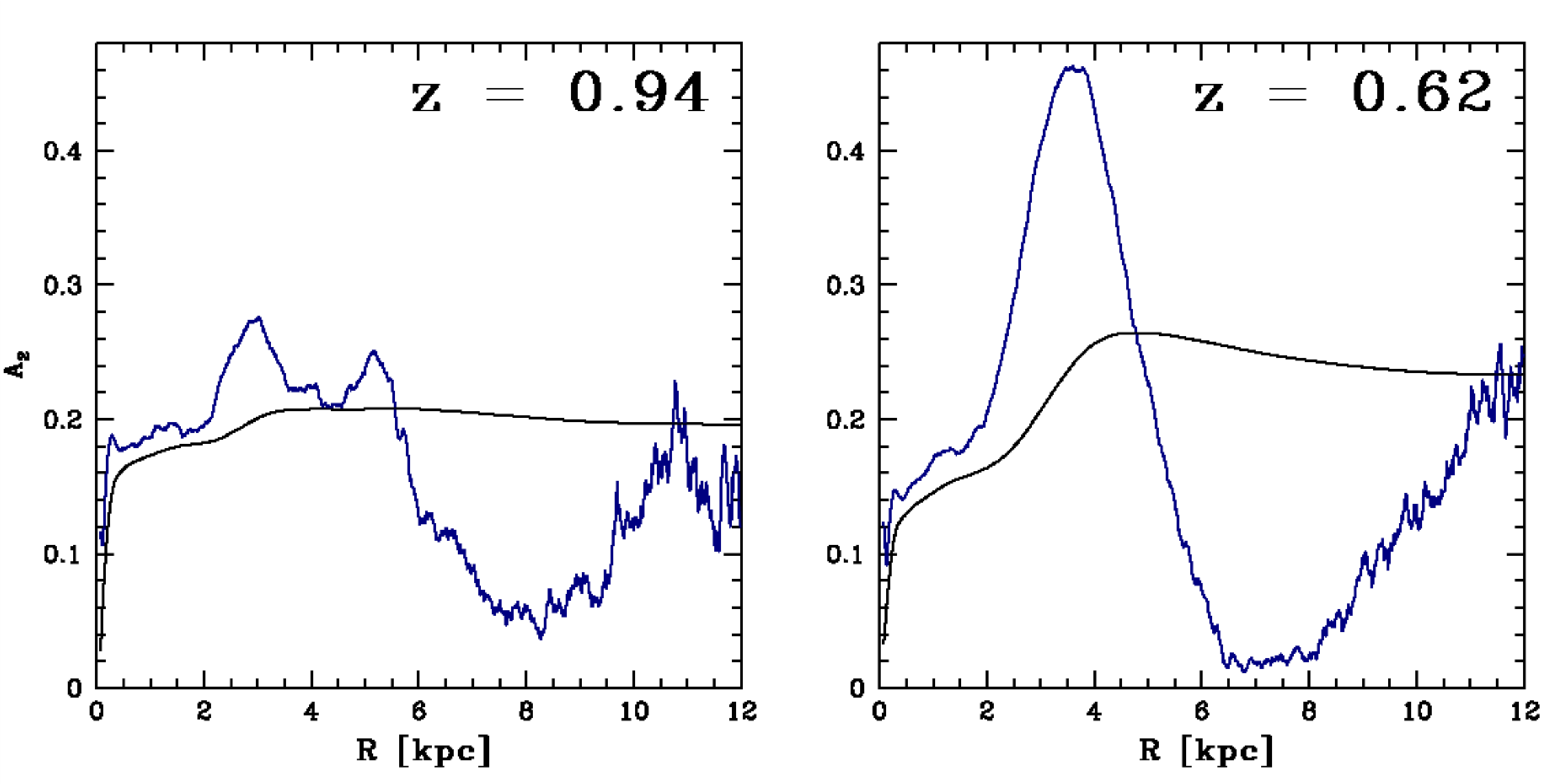}
\caption{Typical outcomes of the Fourier decomposition applied to the Eris2k primary galaxy.
The colour code is the same used in the upper panels of Figure~\ref{fig:a2prof}. The $A_2(R)$ profiles (blue lines) of two non-consecutive snapshots are superimposed on their relative $A_2(<R)$ profiles (black lines).
The presence of various maxima in both the lines of the left panel requires an additional study (see text for more details).}
\label{fig:a2prof_2k}
\end{figure}

As anticipated in Section~\ref{sec:local_scales}, the larger inhomogeneities and the overall higher granularity of the stellar distribution in the Eris2k run, mostly due to the specific feedback prescriptions implemented, result in density profiles less obvious to interpret.
As a consequence, a clear bar is not always evident in every snapshot, for the $A_2(R)$ and the $A_2(<R)$ profiles have likely more than one peak within the investigated radial range.
Figure~\ref{fig:a2prof_2k} shows two typical $A_2$ profiles coming from the Eris2k run.
Differently from the ErisBH case, where almost every profile is unambiguous (as it is exemplified by the three snapshots shown in Figure~\ref{fig:a2prof}), Eris2k offers numerous surface density profiles similar to the one in the left-hand panel of Figure~\ref{fig:a2prof_2k}, where both $A_2(R)$ (blue line) and $A_2(<R)$ (black line) display more crests.

The various peaks could be due, for instance, to a deformation of the bar structure, to the growth of other disc instabilities, or to the presence of a stellar cluster. Thus, the real structure could be linked to any/none (or even more than one) of them.

In order to properly follow the evolution of the two-fold non-axisymmetry, regardless of the environmental disturbances, and to retrieve its correct parameters in the Eris2k main galaxy, we first collect, for each snapshot, a list of all the peaks in the $A_2(<R)$ profile (black lines in Figure~\ref{fig:a2prof_2k}), that correspond to as many $m=2$ overdensities.
Then, for each peak, we check the phase-shift of its corresponding overdensity through Equation~(\ref{eq:phase}), after smoothing the phase at the peak $\Phi(R_{\rm peak})$ with a small kernel in order to obtain a more representative value for that overdensity.
In detail, we select the structure only if the overdensity has constant phase, i.e. if $\left| \Phi(R_{\rm peak})-\Phi(R)\right|<\arcsin(0.15)$, over a radial range $\Delta R > 0.6 \,R_{\rm peak}$.
When more than one overdensity survives this selection, we choose the one with the highest value of $A_2(R)$.

\bsp 
\label{lastpage}
\end{document}